\documentclass[preprint]{aastex}
\usepackage{amstex,amssymb}
\newcommand{\e}[2]{\mbox{$#1 \times 10^{#2}$}}	% Scientific notation
\begin{document}

\title{Properties of Gamma-Ray Burst Time Profiles Using Pulse
Decomposition Analysis}

\author{Andrew Lee and Elliott D. Bloom}
\affil{Stanford Linear Accelerator Center, Stanford University,
Stanford, California 94309}
\makeatletter
\email{alee@slac.stanford.edu and elliott@slac.stanford.edu}
\makeatother

%\and

\author{Vah\'{e} Petrosian}
\affil{Center for Space Science and Astrophysics, Varian 302c,
Stanford University, Stanford, CA 94305-4060 \altaffilmark{1}}
\makeatletter
\email{vahe@astronomy.stanford.edu}
\makeatother

\altaffiltext{1}{Also Astronomy Program and Department of Physics.}

\begin{abstract}
The time profiles of many gamma-ray bursts consist of distinct pulses,
which offers the possibility of characterizing the temporal structure
of these bursts using a relatively small set of pulse shape
parameters.  This pulse decomposition analysis has previously been
performed on a small sample of bright long bursts using binned data
from BATSE, which comes in several data types, and on a sample of
short bursts using the BATSE Time-Tagged Event (TTE) data type.  We
have developed an interactive pulse-fitting program using the
phenomenological pulse model of Norris, \emph{et al.} and a
maximum-likelihood fitting routine.  We have used this program to
analyze the Time-to-Spill (TTS) data for all bursts observed by BATSE
up through trigger number 2000, in all energy channels for which TTS
data is available.  We present statistical information on the
attributes of pulses comprising these bursts, including relations
between pulse characteristics in different energy channels and the
evolution of pulse characteristics through the course of a burst.  We
carry out simulations to determine the biases that our procedures may
introduce.  We find that pulses tend to have shorter rise times than
decay times, and tend to be narrower and peak earlier at higher
energies.  We also find that pulse brightness, pulse width, and pulse
hardness ratios do not evolve monotonically within bursts, but that
the ratios of pulse rise times to decay times tend to decrease with
time within bursts.
\end{abstract}


\section{Introduction}

There has been considerable recent progress in the study of gamma-ray
bursts.  Much of this results from the detection of bursts by BeppoSAX
with good locations that have allowed the detection of counterparts at
other wavelengths.  This has allowed measurements of redshifts that
have firmly established that these bursts are at cosmological
distances.  However, only a few redshifts are known, so there is still
much work to be done in determining the mechanisms that produce
gamma-ray bursts.  Investigation of time profiles and spectra can shed
new light on this subject.

The vast majority of gamma-ray bursts that have been observed have
been observed \emph{only} by BATSE.  This data can be classified into
three major types: burst locations with relatively large
uncertainties, temporal characteristics, and spectral characteristics.
Here, we shall examine temporal characteristics of bursts, along with
some spectral characteristics.

The temporal structure of gamma-ray bursts exhibit very diverse
morphologies, from single simple spikes to extremely complex
structures.  So far, the only clear division of bursts based on
temporal characteristics that has been found is the bimodal
distribution of the $T_{90}$ and $T_{50}$ intervals, which are
measures of burst durations~\citep{kouveliotou:1993,meegan:1996}.  In
order to characterize burst time profiles, it is useful to be able to
describe them using a small number of parameters.

Many burst time profiles appear to be composed of a series of
discrete, often overlapping, pulses, often with a \emph{fast rise,
exponential decay} (FRED) shape~\citep{norris:1996}.  These pulses
have durations ranging from a few milliseconds to several seconds.
The different pulses might, for example, come from different spatial
volumes in or near the burst source.  Therefore, it may be useful to
decompose burst time profiles in terms of individual pulses, each of
which rises from background to a maximum and then decays back to
background levels.  Here, we have analyzed gamma-ray burst time
profiles by representing them in terms of a finite number of pulses,
each of which is described by a small number of parameters.  The BATSE
data used for this purpose is described in Section~\ref{sec:tts}.  The
basic characteristics of the time profiles based on the above model
are described in Section~\ref{sec:results} and some of the
correlations between these characteristics are described in
Section~\ref{sec:correlate}.  (Further analysis of these and other
correlations and their significance are discussed in an accompanying
paper, \cite{lee:2000b}.)  Finally, a brief discussion is presented in
Section~\ref{sec:discuss}.

\section{The BATSE Time-to-Spill Data}
\label{sec:tts}

The BATSE Time-to-Spill (TTS) burst data type records the times
required to accumulate a fixed number of counts, usually 64, in each
of four energy channels~\citep{batse:flight}.  These time intervals
give fixed multiples of the reciprocals of the average count rates
during the spill intervals.  There has been almost no analysis done
using the TTS data because it is less convenient to use with standard
algorithms than the BATSE Time-Tagged Event (TTE) data or the various
forms of binned BATSE data.  The TTS data use the limited memory on
board the CGRO more efficiently than do the binned data types because
at lower count rates, it stores spills less frequently, with each
spill having the same constant fractional statistical error.  On the
other hand, the binned data types always store binned counts at the
same intervals, so that at low count rates the binned counts have a
large fractional statistical error.  The variable time resolution of
the TTS data ranges from under 50~ms at low background rates to under
0.1~ms in the peaks of the brightest bursts.  In contrast, the finest
time resolution available for binned data is 16~ms for the medium
energy resolution (MER) data, and then only for the first 33~seconds
after the burst trigger.  The TTS data can store up to 16,384 spill
events (over $10^{6}$ counts) for each energy channel, and this is
almost always sufficient to record the complete time profiles of
bright, long bursts.  This is unlike the TTE data, which are limited
to 32,768 counts in all four energy channels combined.  For short
bursts, the TTE data have finer time resolution than the TTS data,
because it records the arrival times of individual counts with
2~$\mu$s resolution.  Furthermore, the TTE data also contain data from
before the burst trigger time.  One reason why this is useful is that
some of the shortest bursts are nearly over by the time burst trigger
conditions have been met, so the TTS and MER data aren't very useful
for these bursts.

\begin{figure}\plotone{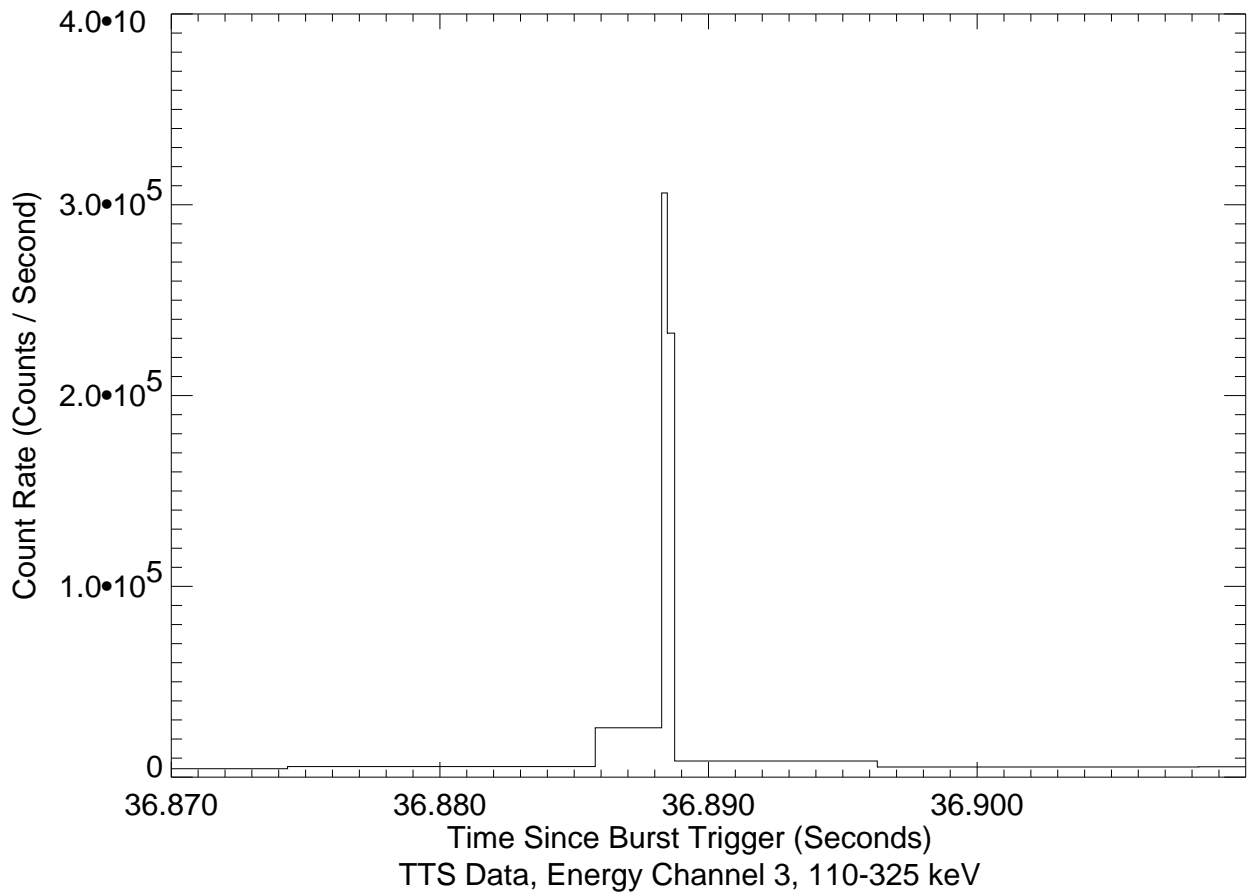}\caption{A portion of the TTS data for BATSE Trigger Number 1577 (4B 920502B).  Note the fine time resolution of the TTS data for the spike more than 36 seconds after the burst trigger.  \label{b01577}}
\end{figure}

Figure~\ref{b01577} shows a portion of the time profile of BATSE
trigger number 1577 (GRB 4B 920502B) that contains a spike with
duration shorter than 1~ms.  The data with the finest time resolution,
the time-tagged event (TTE) data, end long before the spike occurs, so
the TTS data give the best representation of the spike.  The binned
data with the finest time resolution, the MER data with 16~ms bins,
are unavailable for this burst, as are the PREB and DISCSC data with
64~ms bins.

For a Poisson process, the individual event times in the TTE data and
the binned counts in the various binned data types follow the familiar
exponential and the Poisson distributions, respectively.  The spill
times recorded in the TTS data follow the \emph{gamma distribution},
which is the distribution of times needed to accumulate a fixed number
of independent (Poisson) events occurring at a given rate.  The
probability of observing a spill time $t_{s}$ is
\begin{equation}
P(t_{s})\ =\  \frac{t_{s}^{N - 1} R^{N} e^{-Rt_{s}}}{\Gamma(N)}\ ,
\label{eq:gamma}
\end{equation}
where $N$ is the number of events per spill and $R$ is the rate of
individual events.  This probability distribution is closely related
to the Poisson distribution, which gives the number of events
occurring within fixed time intervals for the same process of
independent individual events, such as photon arrivals.

\subsection{The Pulse Model and the Pulse Fitting Procedure}

We now describe the pulse model used to fit GRB time profiles, and the
pulse-fitting procedure.  The pulse model we use is the phenomological
pulse model of \cite{norris:1996} In this model, each pulse is
described by five parameters with the functional form
\begin{equation}
I(t)\  =\  A \exp{\left(-\left\vert\frac{t
 - t_{\text{max}}}{\sigma_{r,d}}\right\vert^{\nu}\right)}\ ,
\label{eq:pulse}
\end{equation}
where $t_{\text{max}}$ is the time at which the pulse attains its
maximum, $\sigma_{r}$ and $\sigma_{d}$ are the rise and decay times,
respectively, $A$ is the pulse amplitude, and $\nu$
(the~``peakedness'') gives the sharpness or smoothness of the pulse at
its peak.  Pulses can, and frequently do, overlap.  \cite{stern:1997b}
have used the same functional form to fit averaged time profiles
(ATPs) of entire bursts.

We have developed an interactive pulse-fitting program that can
automatically find initial background level and pulse parameters using
a Haar wavelet denoised time profile~\citep{donoho:denoise}, and
allows the user to add or delete pulses graphically.  The program then
finds the parameters of the pulses and a background with a constant
slope by using a maximum-likelihood fit for the gamma distribution
(equation~\ref{eq:gamma}) that the TTS spill times
follow~\citep{lee:1996,lee:1998,lee:thesis}.

The data that we use in this paper are the TTS data for all gamma-ray
bursts in the BATSE 3B catalog~\citep{batse:3b} up to trigger number
2000, covering the period from 1991 April 21 through 1992 October 22,
in all channels that are available and show time variation beyond the
normal Poisson noise of the background.  We fit each channel of each
burst separately and obtained 574 fits for 211 bursts, with a total of
2465 pulses.  In many cases, the data for a burst showed no activity
in a particular energy channel, only the normal background counts, so
there were no pulses to fit.  This occurred most frequently in energy
channel~4.  In other cases, the data for a burst contained telemetry
gaps or were completely missing in one or more channels, making it
impossible to fit those channels.

This procedure is likely to introduce selection biases, which can be
quantified through simulation.  To determine these biases, we
simulated a set of bursts with varying numbers of pulses with
distributions of pulse and background parameters based on the observed
distributions in actual bursts.  We generated independent counts
according to the simulated time profiles to create simulated TTS data,
which we subjected to the same pulse-fitting procedure used for the
actual BATSE data.  The detailed results of this simulation are
discussed in the Appendix.  We will contrast the results from the
actual data with those from the simulations where necessary and
relevant.

\subsection{Count Rates and Time Resolution}

The time resolution of the TTS data can be determined from the fitted
background rates and the amplitudes of the individual pulses
(discussed later in Subsection~\ref{sec:amp}), at both the background
levels and at the peaks of the pulses.  Table~\ref{tab:res},
columns~(a) show the percentage of bursts in our fitted sample where
the time resolution at background levels and at the peak of the
highest amplitude pulse are finer than 64~ms and 16~ms, the time
resolutions of the more commonly used DISCSC and MER data,
respectively.  The background rates are taken at the time of the burst
trigger, and ignore the fitted constant slope of the background.  The
rates at the peaks of the highest amplitude pulses include the
background rates at the peak times of the pulses calculated with the
background slopes.  However, these rates ignore overlapping pulses, so
the actual time resolution will be finer since the actual count rates
will be higher.  Note that even at background levels, the TTS data
always have finer time resolution than the DISCSC data, except in
energy channel~4 where the DISCSC data have finer time resolution for
32\% of the bursts in our sample.

\begin{deluxetable}{crrrrrr}
\tablecaption{Percentage of (a) Bursts and (b) Individual Pulses with
Time Resolution $<64$~ms, $<16$~ms. \label{tab:res}} \tablehead{
\colhead{} & \multicolumn{4}{c}{(a) Bursts} & \multicolumn{2}{c}{(b)
Individual} \\ \cline{2-5} \\ \colhead{Energy} &
\multicolumn{2}{c}{Background} & \multicolumn{2}{c}{Highest Amplitude
Pulse} & \multicolumn{2}{c}{Pulses} \\ \colhead{Channel} & \colhead{\%
$<64$~ms} & \colhead{\% $<16$~ms} & \colhead{\% $<64$~ms} &
\colhead{\% $<16$~ms} & \colhead{\% $<64$~ms} & \colhead{\% $<16$~ms}}
\startdata 1 & 100\% & 16\% & 100\% & 80\% & 100\% & 80\% \\ 2 & 100\%
& 8\% & 100\% & 75\% & 100\% & 68\% \\ 3 & 100\% & 3\% & 100\% & 67\%
& 100\% & 62\% \\ 4 & 68\% & 1\% & 100\% & 30\% & 100\% & 42\% \\ All
& 96\% & 8\% & 100\% & 69\% & 100\% & 65\% \\ \enddata
\end{deluxetable}

Table~\ref{tab:res}, columns~(b) show the percentage of individual
pulses where the TTS data have time resolution finer than 16~ms and
64~ms at the pulse peaks.  Again, the count rates include the fitted
background rates at the peak times of the pulses but ignore
overlapping pulses.  For all individual pulses, the TTS data have
finer time resolution at their peaks than the DISCSC data.

\section{General Characteristics of Pulses in Bursts}
\label{sec:results}

In this section we describe characteristics of pulses in individual
bursts and in the sample as a whole.

\subsection{Numbers of Pulses}

\begin{figure}\plotone{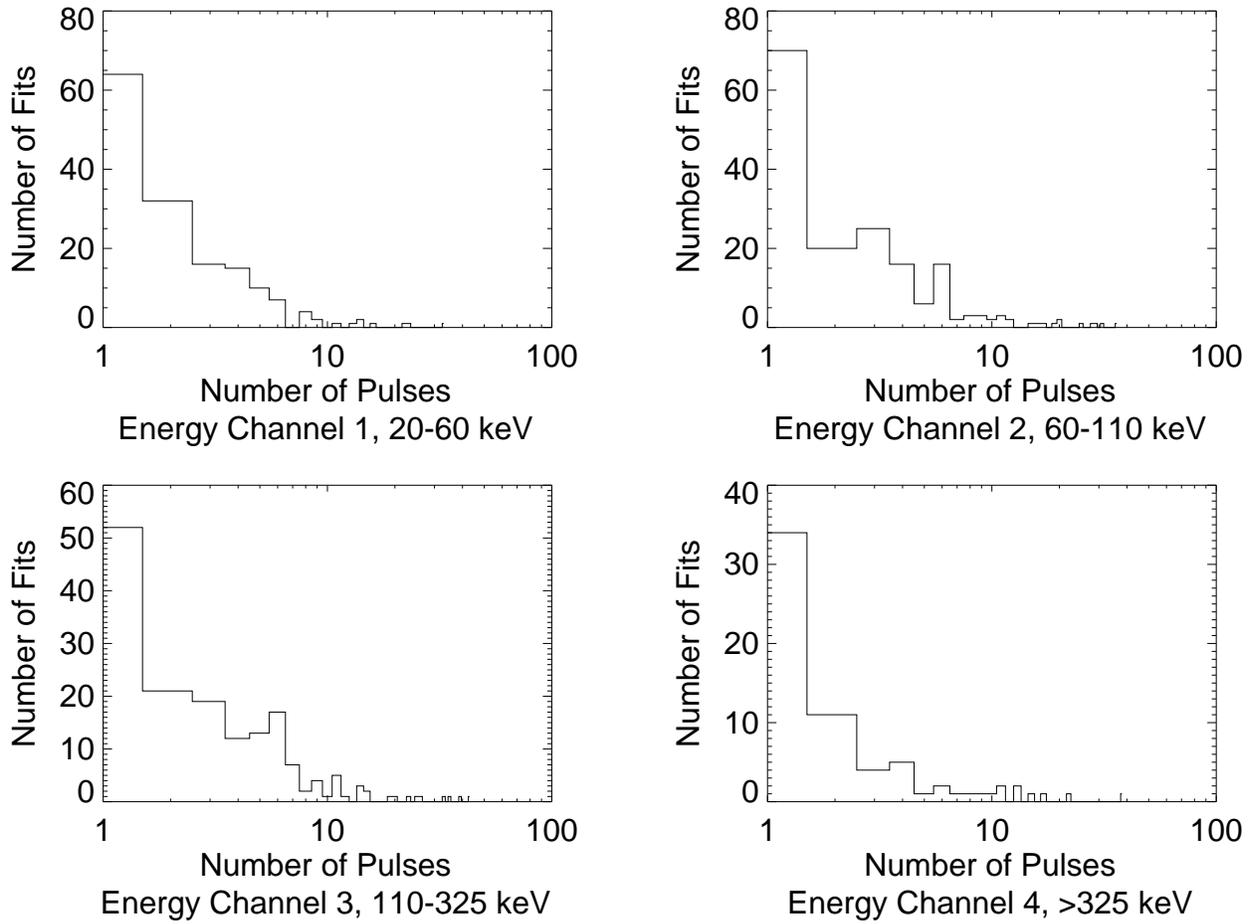}\caption{Distribution of number of pulses per burst, by energy channel.  Compare with Figure~\ref{sim_npulse} from simulations.  \label{npulsech}}
\end{figure}

The number of pulses in a fit range from 1 to 43, with a median of 2
pulses per fit in energy channels~1, 2, and 4, and a median of 3
pulses per fit in energy channel~3.  (See Figure~\ref{npulsech}.)  The
numbers of pulses per fit follows the trend of pulse amplitudes, which
we shall see tend to be highest in energy channel~3, followed in order
by channels~2, 1, and 4, respectively.  This appears to occur because
higher amplitude pulses are easier to identify above the background,
and is consistent with the simulation results shown in the Appendix.

\cite{norris:1996} have used the pulse model of
equation~\ref{eq:pulse} to fit the time profiles of 45 bright, long
bursts.  They analyzed the BATSE PREB and DISCSC data types, which
contain four-channel discriminator data with 64~ms resolution
beginning 2 seconds before the burst trigger.  For their selected
sample of bursts, they fitted an average of 10 pulses per burst, with
no time profiles consisting of only a single pulse.  This number is
considerably higher than the mean number of pulses per fit for our
sample of bursts, probably because their sample was selected for high
peak flux and long duration, which makes it easier to resolve more
pulses.

\subsection{Matching Pulses Between Energy Channels}

To see how attributes of pulses within a burst vary with energy, it is
necessary to match pulses in different energy channels.  Although
burst time profiles generally have similar features in different
energy channels, this matching is not straightforward, since the
number of pulses fitted to a burst time profile is very often
different between energy channels.  We have used a simple automatic
algorithm for matching pulses between adjacent energy channels.  This
algorithm begins by taking all pulses from the channel with fewer
pulses.  It then takes the same number of pulses of highest amplitude
from the other channel, and matches them in time order with the pulses
from the channel with fewer pulses.  For example, the time profiles of
BATSE trigger number 1577 were fitted with nine pulses in energy
channel~3, and only four pulses in channel~4.  This algorithm simply
matches all four pulses in channel~4 in time order with the four
highest amplitude pulses in channel~3.  While this method will not
always correctly match individual pulses between energy channels and
will result in broad statistical distributions, it should still
preserve central tendencies and yield useful statistical information.

\subsection{Brightness Measures of Pulses: Amplitudes and Count Fluences}
%\subsection{Pulse Amplitudes}
\label{sec:amp}

\begin{figure}\plotone{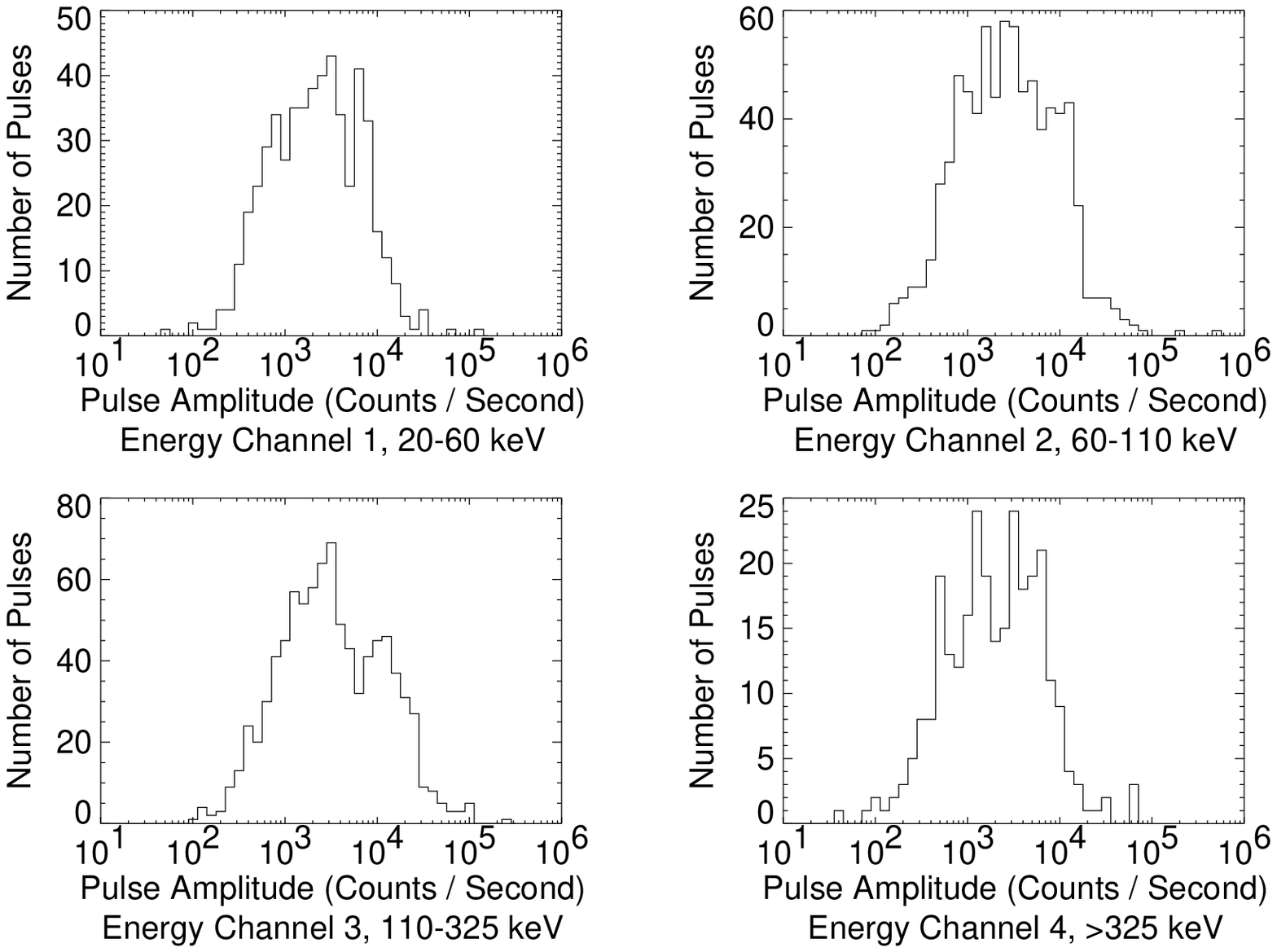}\caption{Distribution of pulse amplitudes for all pulses from all bursts, by energy channel.  Note that the rapid decline at
low amplitudes is partly due to the BATSE triggering procedure and
partly due to the fitting procedure.  See Figure~\ref{sim_amp}.
\label{amp}}
\end{figure}

The amplitude of a pulse, parameter $A$ in equation~\ref{eq:pulse}, is
the maximum count rate within the pulse, and measures the observed
intensity of the pulse, which depends on the absolute intensity of the
pulse at the burst source and the distance to the burst source.  The
amplitudes of the fitted pulses ranged from 40 counts/second to over
500,000 counts/second.  (See Table~\ref{tab:amp} and
Figure~\ref{amp}.)  Pulses tend to have the highest amplitudes in
energy channel~3, followed in order by channels~2, 1, and~4, in
agreement with \cite{norris:1996}.  The central 68\% of the pulse
amplitude distributions span a range of about one order of magnitude
in each of the four energy channels, with a somewhat greater range in
channel~3.  We will see in the Appendix that the fitting procedure
tends to miss pulses with low amplitudes, so that the distributions
shown may be strongly affected by selection effects in the fitting
procedure.

\begin{deluxetable}{crrrr}
\tablecaption{Characteristics of Distribution of Pulse Amplitudes for All Pulses in All Bursts Combined. \label{tab:amp}}
\tablehead{
\colhead{Energy} & \colhead{Min. Amp.} & \colhead{Median Amp.} & \colhead{Max. Amp.} & \colhead{Ratio} \\
\colhead{Channel} & \colhead{(Counts / Sec.)} & \colhead{(Counts / Sec.)} & \colhead{(Counts / Sec.)} & \colhead{84\%ile / 16\%ile}}
\startdata
1 & 47 & 2200 & 136,000 & 10.8 \\
2 & 85 & 2700 & 543,000 & 12.9 \\
3 & 93 & 3000 & 250,000 & 16.2 \\
4 & 43 & 1900 & 63,000 & 11.8 \\
\enddata
\end{deluxetable}

\begin{figure}\plotone{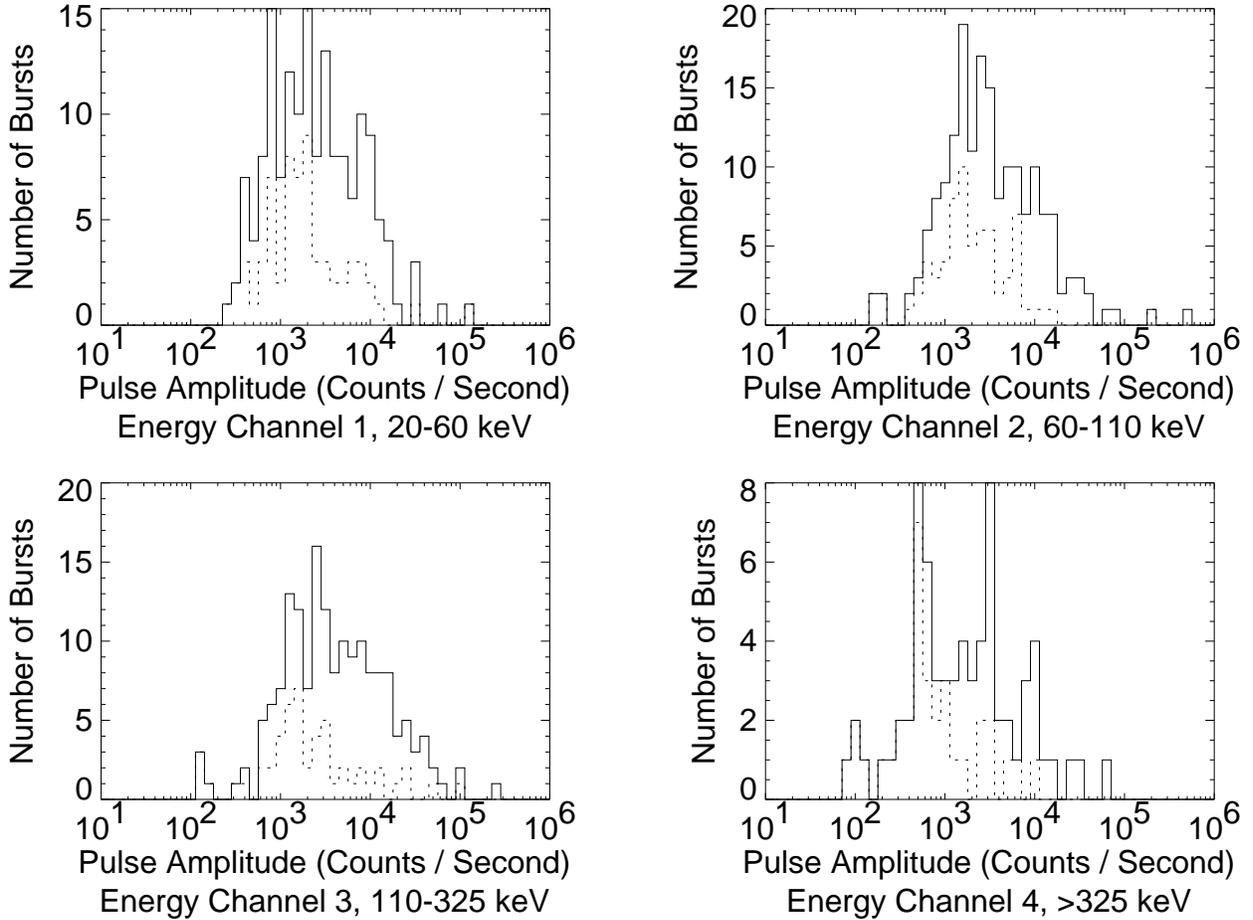}\caption{Distribution of pulse amplitudes, highest amplitude pulse in each burst, by energy channel.  Dashed lines are bursts containing only a single pulse.  The more rapid decline at low
amplitudes as compared with that in Figure~\ref{amp} is due to the
stronger influence of the BATSE triggering procedure.  The fitting
procedure has a weaker influence here.  \label{peak_amp}}
\end{figure}

The amplitude of the highest amplitude pulse in a burst is an
approximation to the instantaneous peak flux above background of that
burst in that energy channel.  The peak flux is often used as an
indicator of the distance to the burst source.  Since pulses can
overlap, the highest pulse amplitude can be less than the actual
background-subtracted peak flux.  The BATSE burst catalogs give
background-subtracted peak fluxes for 64, 256, and 1024~ms time bins
in units of photons/cm$^2$/second, for which effects such as the
energy acceptances of the detectors and the orientation of the
spacecraft and hence the detectors relative to the source have been
accounted for and removed.  The BATSE burst catalog also lists raw
peak count rates that are not background-subtracted or corrected for
any of the effects described, averaged over 64, 256, and 1024~ms time
bins in the second most brightly illuminated detector for each burst.
These peak count rates are primarily useful for comparison with the
BATSE event trigger criteria.  In some bursts, the highest pulses are
considerably narrower than the shortest time bins used to measure peak
flux in the BATSE burst catalog.  For these bursts, these peak fluxes
will be lower than the true peak flux, and the fitted pulse amplitudes
are likely to be a better measure of the true peak flux.  The
distributions of the highest pulse amplitudes are shown in
Table~\ref{tab:peak_amp} and Figure~\ref{peak_amp}.  Since BATSE
selectively triggers on events with high peak flux, the distributions
must be strongly affected by the trigger criteria.

\begin{deluxetable}{crrrr}
\tablecaption{Characteristics of Distribution of Pulse Amplitudes for Highest Amplitude Pulse in Each Burst. \label{tab:peak_amp}}
\tablehead{
\colhead{Energy} & \colhead{Min. Amp.} & \colhead{Median Amp.} & \colhead{Max. Amp.} & \colhead{Ratio} \\
\colhead{Channel} & \colhead{(Counts / Sec.)} & \colhead{(Counts / Sec.)} & \colhead{(Counts / Sec.)} & \colhead{84\%ile / 16\%ile}}
\startdata
1 & 241 & 2200 & 136,000 & 11.1 \\
2 & 148 & 2800 & 543,000 & 9.9 \\
3 & 116 & 3500 & 250,000 & 12.4 \\
4 & 82 & 1500 & 63,000 & 18.8 \\
\enddata
\end{deluxetable}

\begin{figure}\plotone{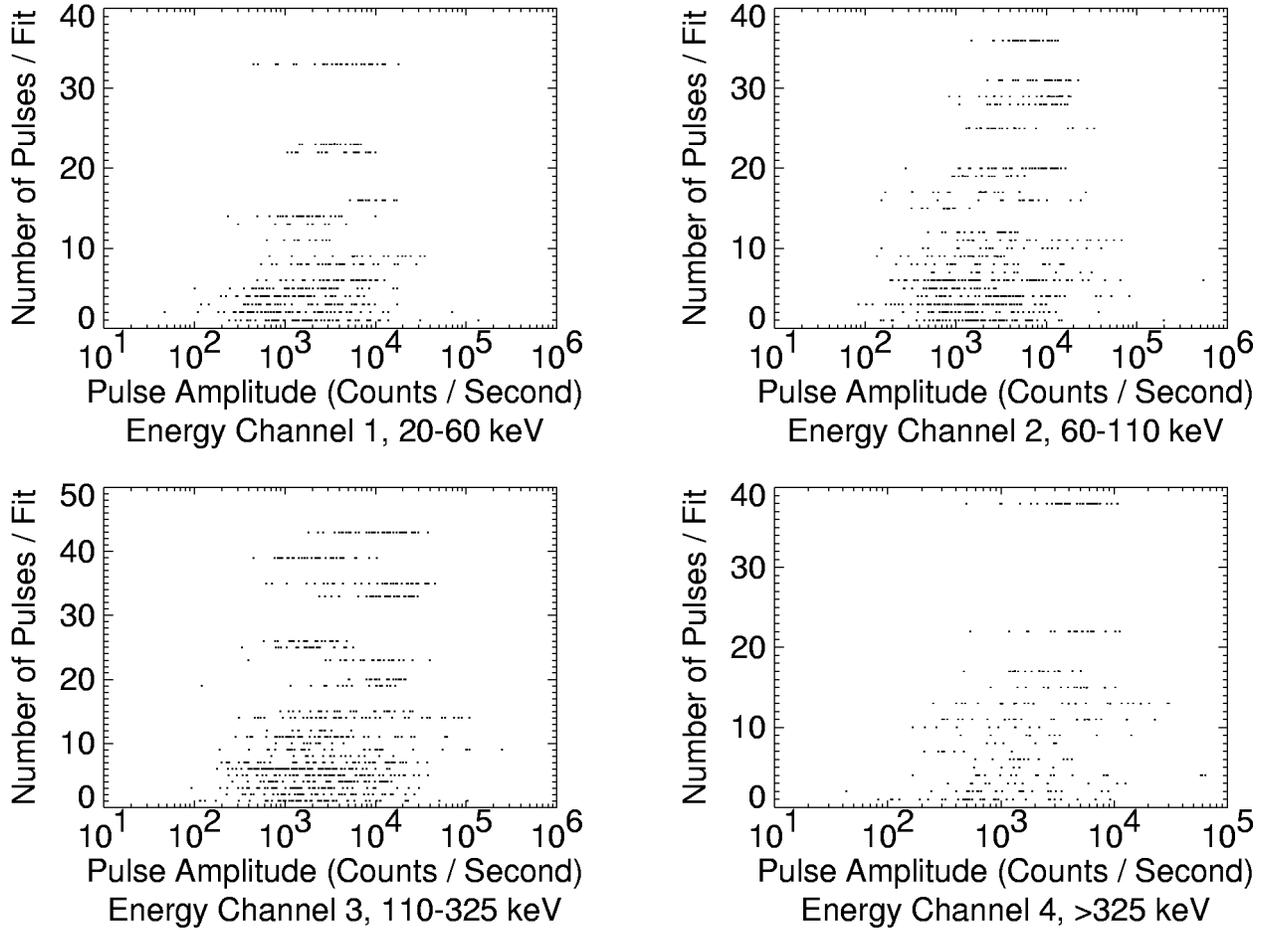}\caption{Number of pulses per burst versus pulse amplitudes of all pulses, by energy channel.  Compare with Figure~\ref{sim_npvsamp} for simulated data.  Note that there exists a positive
correlation between the two quantities.  \label{npvsamp}}
\end{figure}

Figure~\ref{npvsamp} shows the number of pulses in each fit plotted
against the amplitudes of all of the pulses comprising each fit.  It
shows that in fits with more pulses, the minimum pulse amplitude,
which can be seen from the left boundary of the distribution, tends to
be higher.  This could result in part from intrinsic properties of the
burst sources, but may also result at least in part from a selection
effect: In a complex time profile with many overlapping pulses, low
amplitude pulses, which have poor signal-to-noise ratios, will be more
difficult to resolve, while in a less complex time profile, they will
be easier to resolve.  This hypothesis appears to be confirmed by the
simulation results shown in the Appendix.  Table~\ref{tab:npvsafw},
columns~(a) give the Spearman rank-order correlation coefficients,
commonly denoted as $r_{s}$, for the joint distribution of pulse
amplitudes and numbers of pulses in the corresponding bursts shown in
Figure~\ref{npvsamp}, as well as the probability that a random data
set of the same size with no correlation between the two variables
would produce the observed value of $r_{s}$.  It shows strong positive
correlations between pulse amplitudes and the number of pulses in the
fit for all energy channels.  These correlations appear to be stronger
than those arising in the fits to simulations shown in
Table~\ref{tab:sim_npvsafw}, columns~(a).

\begin{deluxetable}{crlrlrl}
\tablecaption{Correlation Between Number of Pulses per Burst and (a) Amplitudes, (b) Count Fluences, and (c) Widths of All Pulses. \label{tab:npvsafw}}
\tablehead{
\colhead{Energy} & \multicolumn{2}{c}{(a) Amplitude} & \multicolumn{2}{c}{(b) Count Fluence} & \multicolumn{2}{c}{(c) Width} \\
\colhead{Channel} & \colhead{$r_{s}$} & \colhead{Prob.} & \colhead{$r_{s}$} & \colhead{Prob.} & \colhead{$r_{s}$} & \colhead{Prob.}}
\startdata
1 & 0.37 & \e{3.9}{-18} & -0.17 & \e{9.2}{-5} & -0.36 & \e{1.8}{-17} \\
2 & 0.36 & \e{1.8}{-24} & -0.20 & \e{3.4}{-8} & -0.35 & \e{8.0}{-24} \\
3 & 0.30 & \e{4.2}{-20} & -0.04 & 0.21 & -0.23 & \e{9.4}{-12} \\
4 & 0.45 & \e{2.2}{-15} & -0.13 & 0.027 & -0.28 & \e{1.3}{-6} \\
\enddata
\end{deluxetable}

%\subsection{Count Fluences}

The area under the light curve of a pulse gives the total number of
counts contained in the pulse, which is its count fluence.  It is
given in terms of the pulse parameters and the gamma function by
\begin{equation}
\mathcal{F} = A \int_{-\infty}^{\infty}{I(t)dt} = A \frac{\sigma_{r} + \sigma_{d}}{\nu}\Gamma\left(\frac{1}{\nu}\right) .
\label{eq:area}
\end{equation}
The count fluence is a measure of the observed integrated luminosity
of the pulse, which depends on the total number of photons emitted by
the source within the pulse and the distance to the burst source.  We
will see in the Appendix that the fitting procedure tends to miss
pulses with low count fluences.

\begin{figure}\plotone{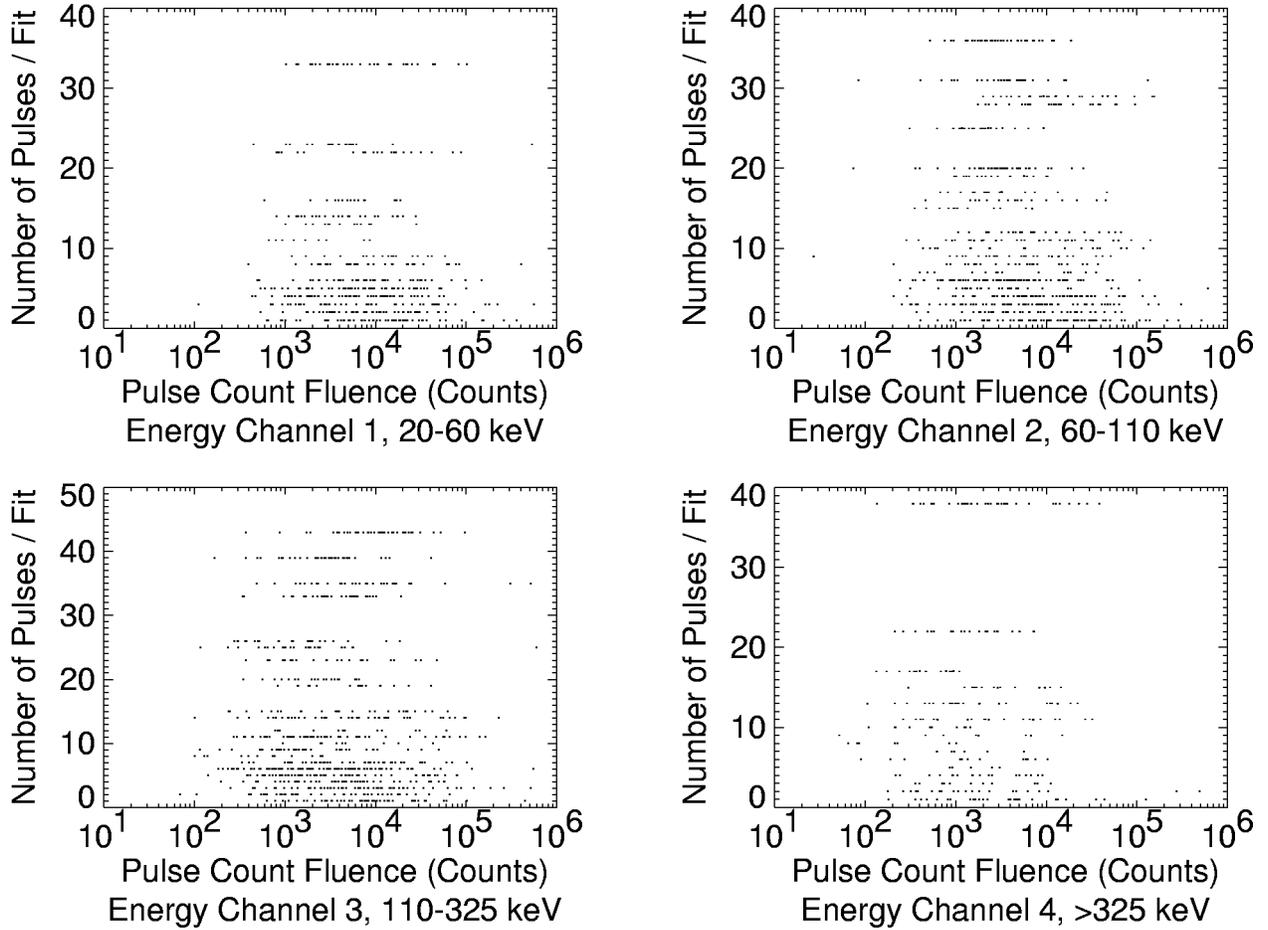}\caption{Number of pulses per burst versus count fluences of all pulses from all bursts, by energy channel.  Compare with Figure~\ref{sim_npvsarea} for simulated data.  Note that there exists a negative correlation between the two quantities.  \label{npvsarea}}
\end{figure}

Figure~\ref{npvsarea} shows the number of pulses in each fit versus
the count fluences of the individual pulses.  It shows that in bursts
containing more pulses, the individual pulses tend to contain fewer
counts.  We shall see in the next section that pulses tend to be
narrower in more complex bursts.  This result for count fluences
implies that the tendency for pulses to be narrower is stronger than
the tendency for pulses to have higher amplitudes in more complex
bursts.  Table~\ref{tab:npvsafw}, columns~(b) show that the
corresponding negative correlations between pulse count fluences and
numbers of pulses per fit are statistically significant in energy
channels~1 and 2, but not in channels~3 and 4.  The fits to
simulations (Figure~\ref{sim_npvsarea} and
Table~\ref{tab:sim_npvsafw}, columns~(b)) do not show the same
tendency, so this most likely is not caused by selection effects in
the pulse-fitting procedure.

%\subsection{Timescales in Bursts}
\subsection{Pulse Widths and Time Delays}

Timescales in gamma-ray bursts are likely to be characteristic of the
physical processes that produce them.  However, since some, and
possibly all, bursts are produced at cosmological distances, all
observed timescales will be affected by cosmological time dilation,
and won't represent the physical timescales at the sources.

\subsubsection{Pulse Widths}

\begin{figure}\plotone{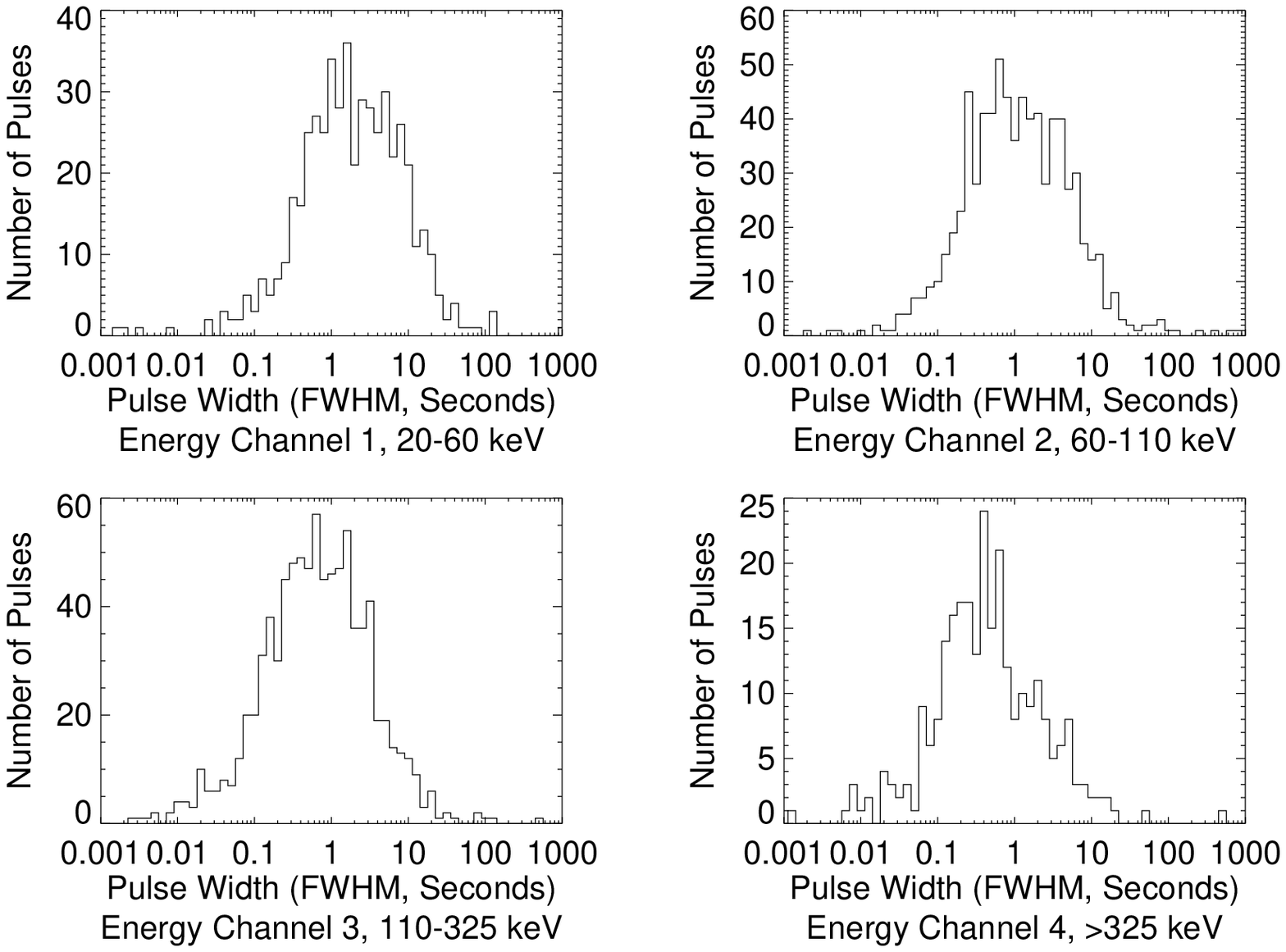}\caption{Distribution of pulse widths (FWHM) for all pulses from all bursts, by energy channel.  Note that there is no indication that pulse widths have the bimodality observed in the distributions of the $T_{50}$ and $T_{90}$ measures of burst durations in the BATSE catalogs.  Compare with the distribution for simulated
bursts in Figure~\ref{sim_fwhm}.  \label{fwhmch}}
\end{figure}

The most obvious timescale that appears in the pulse decomposition of
gamma-ray burst time profiles is the pulse width, or duration.  We
shall measure the duration, or width, of a pulse using its full width
at half maximum (FWHM), which is given by
\begin{equation}
T_{\text{FWHM}} = (\sigma_{r} + \sigma_{d}) (\ln 2)^{\frac{1}{\nu}} .
\label{eq:fwhm}
\end{equation}
The distributions of the pulse widths, which are shown in
Figure~\ref{fwhmch} and columns~(a) of Table~\ref{tab:fwhmnu}, peak
near one second in all energy channels, with no sign of the bimodality
seen in total burst durations mentioned above.  Pulses tend to be
narrower (shorter) at higher energies.

\begin{deluxetable}{crrrr}
\tablecaption{Characteristics of Distribution of (a) Pulse Widths and of (b) Peakedness $\nu$ for All Pulses in All Bursts Combined. \label{tab:fwhmnu}}
\tablehead{
\colhead{} & \multicolumn{2}{c}{(a) FWHM} & \multicolumn{2}{c}{(b) Peakedness $\nu$} \\
\colhead{Energy} & \colhead{Median} & \colhead{Ratio} & \colhead{Median} & \colhead{Ratio} \\
\colhead{Channel} & \colhead{(Seconds)} & \colhead{84\%ile / 16\%ile} & \colhead{$\nu$} & \colhead{84\%ile / 16\%ile}}
\startdata
1 & 1.86 & 19.9 & 1.22 & 5.3 \\
2 & 1.05 & 23.0 & 1.26 & 5.6 \\
3 & 0.68 & 22.9 & 1.26 & 5.7 \\
4 & 0.41 & 21.4 & 1.17 & 5.8 \\
All & 0.90 & 27.0 & 1.25 & 5.6 \\
\enddata
\end{deluxetable}

The narrowing of pulses in higher energy channels can also be measured
from the ratios of pulse widths of matched pulses in adjacent energy
channels, as shown in Table~\ref{tab:rfwhm}.  We can test the
hypothesis that pulses tend to be narrower at higher energies by
computing the probability that the observed numbers of pulses width
ratios less than 1 will occur by chance if pulse width ratios less
than 1 and greater than 1 are equally probable.  This probability can
be computed from the binomial distribution, and is shown in the last
column of Table~\ref{tab:rfwhm}.  The table shows less narrowing than
a simple comparison of median pulse widths from Table~\ref{tab:fwhmnu}
would suggest, though it also shows that the hypothesis that pulses
\emph{do not} become narrower at higher energies is strongly excluded
between channels~1 and 2 and between channels~2 and 3.  Qualitatively
similar kinds of trends have been shown to be present in individual
pulses~\citep{norris:1996} and composite pulse shapes of many
bursts~\citep{link:1993,fenimore:1995b}.  There are, however, some
quantitative differences.  For example, we find that there seems to be
less narrowing at higher energies; the pulse width ratios tend to be
closer to 1 between energy channels~3 and 4 than for the lower energy
channels (although the statistics are poorer, as with anything
involving channel~4), which is the opposite of the tendency found by
\cite{norris:1996}.  We can use the Kolmogorov-Smirnov test to
determine if the distributions of pulse width ratios are the same
between adjacent energy channels.  These results are shown in the last
column of Table~\ref{tab:rfwhm}.  This test shows significant
differences in the distribution of pulse widths of matched pulses
between adjacent energy channels.

\begin{deluxetable}{crrll}
\tablecaption{Ratios of Pulse Widths of Pulses Matched Between Adjacent Energy Channels. \label{tab:rfwhm}}
\tablehead{
\colhead{Energy} & \colhead{Median} & \colhead{} & \colhead{Binom.} & \colhead{K-S} \\
\colhead{Channels} & \colhead{Width Ratio} & \colhead{\% $<1$} & \colhead{Prob.} & \colhead{Prob.}}
\startdata
2 / 1 & 0.73 & 304/446 = 68\% & \e{1.7}{-14} & \e{1.1}{-5} \\
3 / 2 & 0.68 & 436/625 = 70\% & $<10^{-16}$ & \e{8.7}{-7} \\
4 / 3 & 0.83 & 153/258 = 59\% & 0.0028 & 0.62 \\
\enddata
\end{deluxetable}

The fact that pulse widths decrease monotonically with energy, and the
signal-to-noise ratios of the different energy channels increase in
order of the energy channels 3, 2, 1, 4, imply that the narrowing is
caused by the burst production mechanism itself.

\subsubsection{Pulse Widths and Numbers of Pulses}

\begin{figure}\plotone{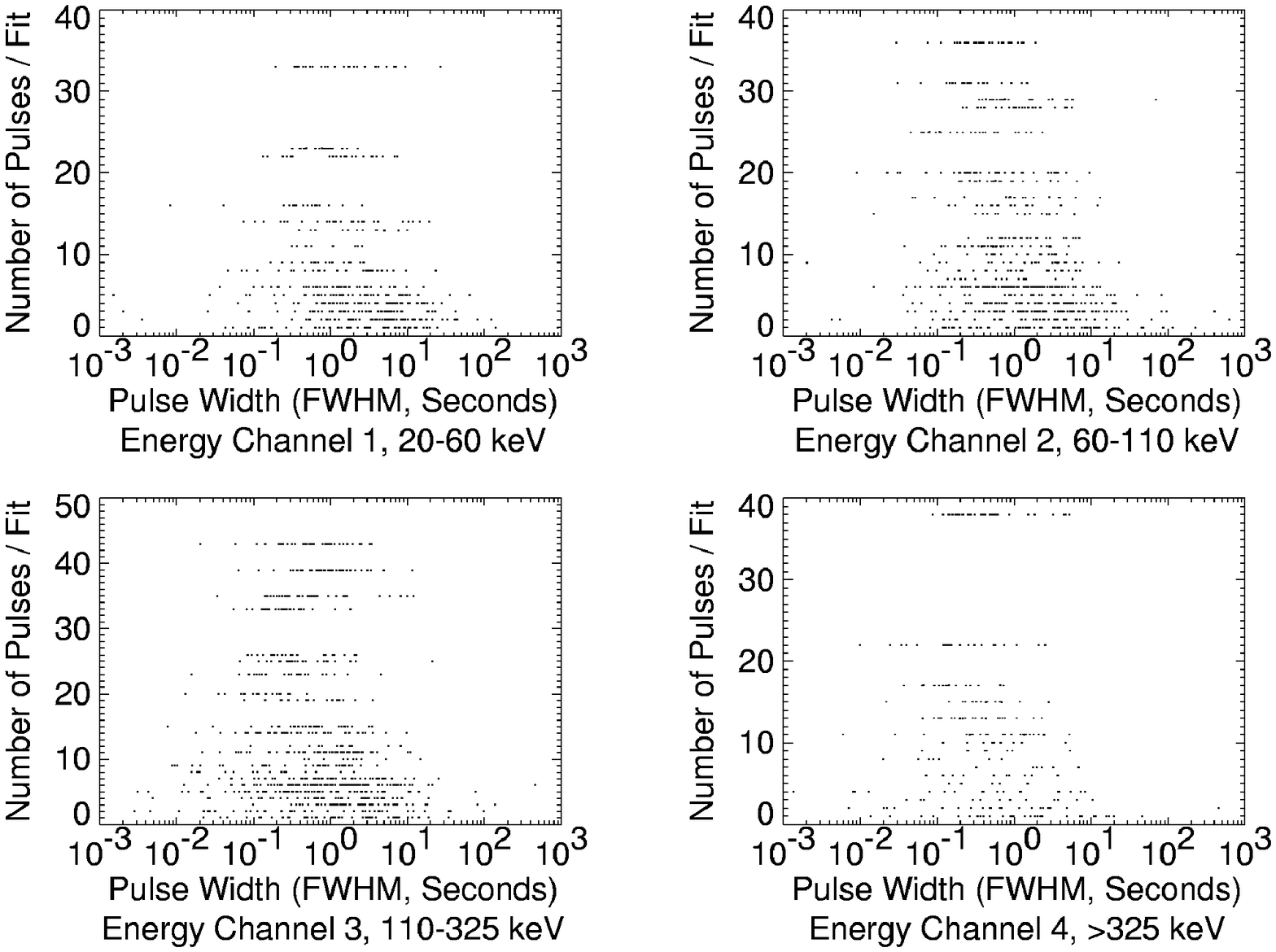}\caption{Number of pulses per burst versus widths of all pulses from all bursts, by energy channel.  Compare with Figure~\ref{sim_npvsfwhm} for simulated data.  Note that there exists a negative correlation
between the two quantities.  \label{npvsfwhm}}
\end{figure}

Figure~\ref{npvsfwhm} and Table~\ref{tab:npvsafw}, columns~(c) show
the relation between the number of pulses per burst and the widths of
the pulses.  These show that pulses tend to be narrower in bursts with
more pulses.  This may be an intrinsic property of GRBs, or it may be
a selection effect arising because narrower pulses have less overlap
with adjacent pulses, hence they are easier to resolve, so more pulses
tend to be identified in bursts with narrower pulses.  This may also
be a side effect of correlations between other burst and pulse
characteristics with the number of pulses per burst and the pulse
widths.  Table~\ref{tab:npvsafw} shows strong negative correlations
between the numbers of pulses per fit and the pulse widths.  The fits
to simulations shown in Figure~\ref{sim_npvsfwhm} and
Table~\ref{tab:sim_npvsafw}, columns~(c) do not have the same
tendency.  This suggests that the negative correlation between the
number of pulses in each fit and the pulse widths seen in the fits to
actual bursts do not result from selection effects in the
pulse-fitting procedure, but are intrinsic to the burst production
mechanism, or may arise from other effects.

\subsubsection{Time Delays Between Energy Channels}
\label{sec:delay}

Table~\ref{tab:timelag}, columns~(a) show the differences, or time
delays, between the peak times $t_{\text{max}}$ of all pulses matched
between adjacent energy channels.  It shows a significant tendency for
individual pulses to peak earlier at higher energies.  This has been
previously observed, and described as a hard-to-soft spectral
evolution of the individual pulses~\citep{norris:1986,norris:1996}.
The time delays found here are greater than those found by
\cite{norris:1996}, who found an average pulse peak time delay between
adjacent energy channels of $\sim 20$ ms.  Comparing the peak times of
the highest amplitude pulses in each fit between adjacent energy
channels also shows a significant tendency for bursts to peak earlier
at higher energies.  (See Table~\ref{tab:timelag}, columns~(b).)  The
time delays between energy channels observed here and elsewhere are
likely to result from intrinsic properties of the burst sources.

\begin{deluxetable}{crrlrrl}
\tablecaption{Characteristics of Distribution of Time Delays Between Adjacent Energy Channels. \label{tab:timelag}}
\tablehead{
\colhead{} & \multicolumn{3}{c}{(a) All Matched Pulses} & \multicolumn{3}{c}{(b) Highest Amplitude Pulse} \\
\colhead{Energy} & \colhead{Med. Lag} & \colhead{} & \colhead{Binom.} & \colhead{Med. Lag} & \colhead{} & \colhead{Binom.} \\
\colhead{Channels} & \colhead{(Sec.)} & \colhead{\% $>0$} & \colhead{Prob.} & \colhead{(Sec.)} & \colhead{\% $>0$} & \colhead{Prob.}}
\startdata
1 - 2 & 0.11 & 290/446 = 65\% & \e{2.2}{-10} & 0.08 & 95/141 = 67\% & \e{3.7}{-5} \\
2 - 3 & 0.27 & 459/625 = 73\% & $<10^{-16}$ & 0.05 & 97/151 = 64\% & 0.00047 \\
3 - 4 & 0.01 & 140/258 = 54\% & 0.17 & 0.14 & 47/67 = 70\% & 0.00097 \\
\enddata
\end{deluxetable}

\subsection{Pulse Shapes: Asymmetries and the Peakedness $\nu$}
%\subsection{Pulse Asymmetries}

\begin{figure}\plotone{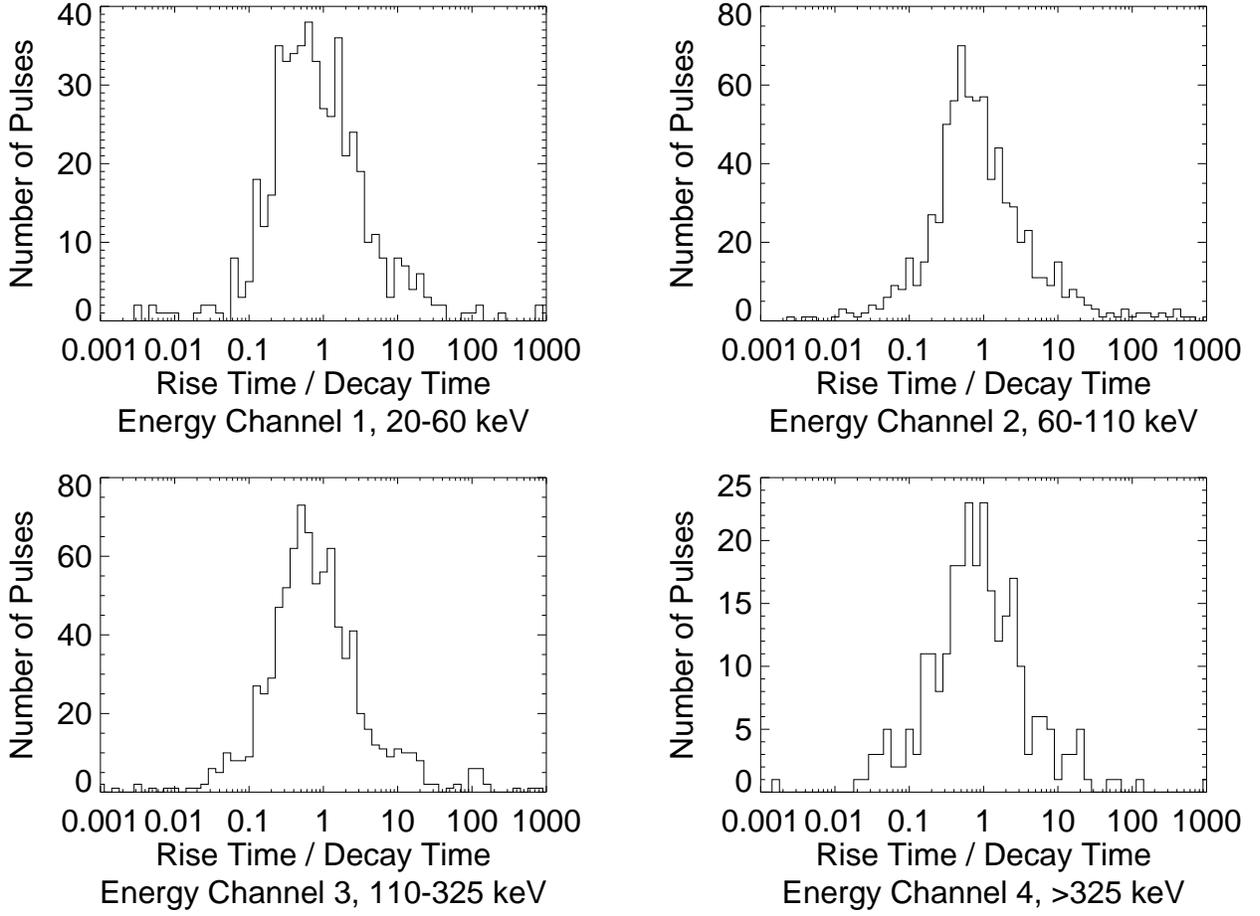}\caption{Distribution of pulse asymmetry ratios for all pulses from all bursts, by energy channel.  See analysis in Table~\ref{tab:rd}.  \label{rdch}}
\end{figure}

Although the pulse model uses separate rise and decay times as its
basic parameters, it is often more natural to consider the widths and
asymmetries of pulses, which give equivalent information to the rise
and decay times.  The ratios of pulse rise times to decay times
$\sigma_{r}/\sigma_{d}$ are a convenient way to measure the asymmetry
of pulses, and depends only on the shapes of pulses.  The asymmetry
ratios cover a very wide range of values, but there is a clear
tendency for pulses to have slightly shorter rise times than decay
times.  (See Figure~\ref{rdch}.)

Table~\ref{tab:rd} shows that the hypothesis that pulses are symmetric
is strongly excluded in energy channels~2 and 3.  The binomial
probability isn't computed for all pulses in all energy channels
combined, because pulses cannot be considered to be independent
between energy channels.  We also see that the degree of the asymmetry
isn't significantly different for the different energy channels.
\cite{norris:1996} found far greater asymmetry, with average values of
$\sigma_{d}/\sigma_{r}$ (the inverse of the ratio used here) ranging
from 2 to 3 for their selected sample of bursts, and with about 90\%
of pulses having shorter rise times than decay times.

\begin{deluxetable}{crrrl}
\tablecaption{Characteristics of Distribution of Pulse Asymmetries for All Pulses in All Bursts Combined. \label{tab:rd}}
\tablehead{
\colhead{Energy} & \colhead{Median} & \colhead{Ratio} & \colhead{} & \colhead{Binom.} \\
\colhead{Channel} & \colhead{$\sigma_{r} / \sigma_{d}$} & \colhead{84\%ile / 16\%ile} & \colhead{\% $\sigma_{r} < \sigma_{d}$} & \colhead{Prob.}}
\startdata
1 & 0.76 & 16.8 & 297/526 = 56\% & 0.0030 \\
2 & 0.76 & 16.1 & 457/776 = 59\% & \e{7.2}{-7} \\
3 & 0.71 & 14.0 & 528/883 = 60\% & \e{5.8}{-9} \\
4 & 0.80 & 15.9 & 158/280 = 56\% & 0.031 \\
All & 0.75 & 15.4 & 1440/2465 = 58\% & \nodata \\
\enddata
\end{deluxetable}

%\subsection{The Peakedness Parameter $\nu$}

\begin{figure}\plotone{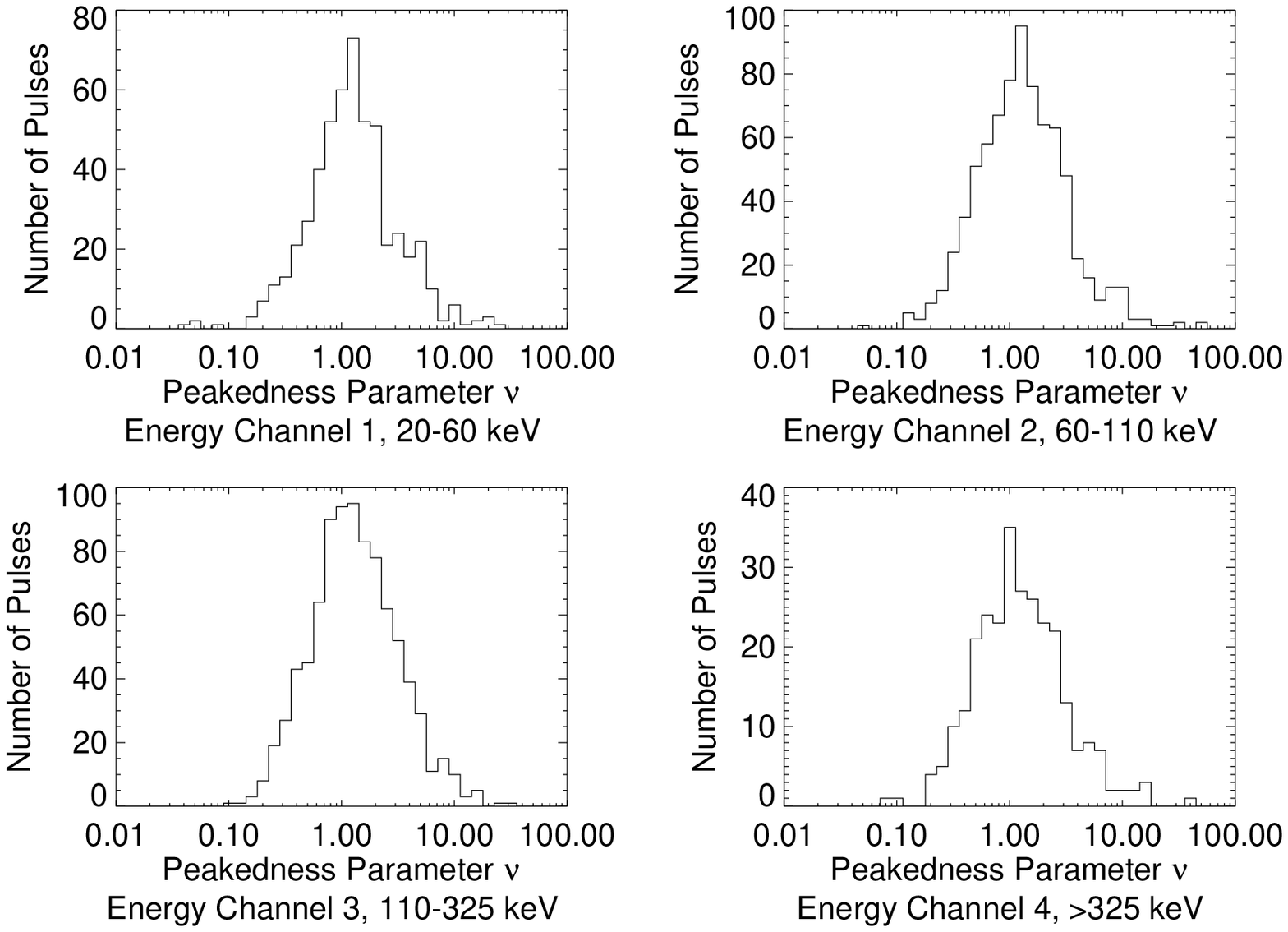}\caption{Distribution of the peakedness parameter $\nu$ for all pulses from all bursts, by energy channel.  See analysis in Table~\ref{tab:fwhmnu}, columns~(b).  \label{nuch}}
\end{figure}

The relation of the peakedness parameter $\nu$ to physical
characteristics of gamma-ray burst sources is far less clear than for
other pulse attributes.  Nevertheless, it does give information that
can be used to compare the shapes of different pulses.  The peakedness
$\nu$ has a median value near 1.2 in all energy channels, so that
pulses tend to have shapes between an exponential, for which $\nu =
1$, and a Gaussian, for which $\nu = 2$.  (See Figure~\ref{nuch} and
columns~(b) of Table~\ref{tab:fwhmnu}.)  \cite{stern:1997b} use the
functional form of equation~\ref{eq:pulse} to fit averaged time
profiles of many bursts rather than individual constituent pulses, and
find that $\nu \approx 1/3$ for the \emph{averaged time profiles}.

\section{Correlations Between Pulse Characteristics}
\label{sec:correlate}

Correlations between different characteristics of pulses, or the lack
thereof, may reveal much about gamma-ray bursts that the distributions
of the individual characteristics cannot.  Some correlations may arise
from intrinsic properties of the burst sources, while others may
result from the differing distances to the sources.  The first kind of
correlation may be present among pulses of individual bursts or among
the whole population of bursts, while the second kind will not be
present among pulses of individual bursts.  In order to distinguish
between these two kinds of effects, it is useful to examine
correlations of pulse characteristics both between different bursts,
and between pulses within individual bursts.

It is simplest to find correlations between characteristics of all
pulses, but such correlations would combine both kinds of effects, and
the statistics would be weighted in favor of bursts containing more
pulses.  It is also possible to select a single pulse from each burst,
and find correlations between the characteristics of these pulses from
burst to burst in order to look for effects arising from the distances
to burst sources.  However, if the correlations are taken using the
single highest amplitude or highest fluence pulse from each burst,
then they could still be affected by correlations of pulse
characteristics within individual bursts.  For example, consider a
situation where amplitudes and durations of pulses within individual
bursts are correlated, and where pulse amplitudes and durations follow
a common distribution for all bursts.  In such a case, if we select
the single highest amplitude pulse from each burst, we would find a
spurious correlation between highest pulse amplitude and duration
between different bursts.

Correlation results which compare and contrast the cosmological and
intrinsic effects will be discussed in greater detail in the
accompanying paper \cite{lee:2000b}.  Here, we describe our method and
some other correlation results.

One way to find correlations of pulse characteristics within
individual bursts is to calculate a correlation coefficient for each
burst and examine the distribution of the degrees of correlation, for
example to see if the correlation coefficients were positive for a
large majority of bursts.  The Spearman rank-order correlation
coefficient is used for this purpose here.  When using the Spearman
rank-order correlation coefficient, the coefficients for the
individual bursts are often not statistically significant because the
number of pulses in each burst is not large, even though the
coefficents for the different bursts may be mostly positive or mostly
negative.  We can test the hypothesis that there is no correlation
because in the absence of any correlation, we would expect an equal
number of bursts with positive correlations as with negative
correlations, so the probability that the observed numbers of bursts
with positive and negative correlations could occur by chance if there
was no correlation is given by the binomial distribution.  This is the
method used here.  This method ignores the strengths of the individual
correlations, so it is more sensitive to a weak correlation that
affects large numbers of bursts than it is to a strong correlation
that affects only a small number of bursts.

\subsection{Spectral Characteristics}

The data that we are using only has very limited spectral information,
only four energy channels.  We can investigate spectral
characteristics by using the \emph{hardness ratios} of individual
pulses.  The hardness ratio of a pulse between two specified energy
channels is the ratio of the fluxes or fluences of the pulse between
the two energy channels.  Although the actual numerical values of the
hardness ratios depend on the somewhat arbitrary boundaries of the
energy channels, the values can be compared between different pulses,
and between different bursts.

There have been several claims of correlation between peak or average
hardness ratios and durations among bursts, with shorter bursts being
harder, and there has been some analysis of the cosmological
significance of this.  Here we investigate similar correlations for
bursts, and for pulses in individual bursts.

\subsubsection{Pulse Widths}

Table~\ref{tab:hrampvswdta}, columns~(a) show the correlations between
the pulse amplitude hardness ratios and the pulse widths for the
highest amplitude pulse in each burst.  The pulse widths used are
arithmetic means of the widths in the two adjacent energy channels
that the hardness ratios are taken between, \emph{e.g.}  hardness
ratios between channels~2 and 3 are compared with pulse widths
averaged over channels~2 and 3.  In all pairs of adjacent energy
channels, the highest amplitude pulse has a slight tendency to be
narrower when the burst is harder, as measured using peak flux, but
this does not appear to be statistically significant, except possibly
between channels~3 and 4.  This may be a signature of weak redshift
effects; whereby the higher the redshift, the softer the spectrum and
the longer the duration.

\begin{deluxetable}{crlrlrl}
\tablecaption{Correlations Between Pulse Amplitude Hardness Ratio and (a) Pulse Widths, (b) Intervals Between Two Highest Amplitude Pulses, and (c) Amplitudes for Highest Amplitude Pulse(s) in Each Burst. \label{tab:hrampvswdta}}
\tablehead{
\colhead{Energy} & \multicolumn{2}{c}{(a) Width} & \multicolumn{2}{c}{(b) Interval} & \multicolumn{2}{c}{(c) Amplitude} \\
\colhead{Channels} & \colhead{$r_{s}$} & \colhead{Prob.} & \colhead{$r_{s}$} & \colhead{Prob.} & \colhead{$r_{s}$} & \colhead{Prob.}}
\startdata
2 / 1 & -0.11 & 0.18 & -0.26 & 0.019 & 0.01 & 0.89 \\
3 / 2 & -0.21 & 0.010 & -0.06 & 0.57 & 0.28 & 0.00059 \\
4 / 3 & -0.41 & 0.00061 & -0.17 & 0.34 & 0.31 & 0.012 \\
\enddata
\end{deluxetable}

Table~\ref{tab:hrampfwhmamp}, columns~(a) shows the correlations
between pulse amplitude hardness ratios and pulse widths within
bursts.  As evident, there is almost equal probability for positive
and negative correlations.  We conclude, therefore, that there is no
significant tendency for longer or shorter duration pulses to have
harder or softer spectra, measured using peak flux.

\begin{deluxetable}{crlrl}
\tablecaption{Correlations Between Pulse Amplitude Hardness Ratio and (a) Pulse Width and (b) Amplitude Within Bursts. \label{tab:hrampfwhmamp}}
\tablehead{
\colhead{Energy} & \multicolumn{2}{c}{(a) Width} & \multicolumn{2}{c}{(b) Amplitude} \\
\colhead{Channels} & \colhead{\% Pos. Corr.} & \colhead{Binom. Prob.} & \colhead{\% Pos. Corr.} & \colhead{Binom. Prob.}}
\startdata
2 / 1 & 45/83 = 54\% & 0.44 & 44.5/83 = 54\% & 0.51 \\
3 / 2 & 50/95 = 53\% & 0.61 & 49/95 = 52\% & 0.76 \\
4 / 3 & 14/33 = 42\% & 0.38 & 20.5/33 = 62\% & 0.16 \\
All & 109/211 = 52\% & \nodata & 114/211 = 54\% & \nodata \\
\enddata
\end{deluxetable}

\subsubsection{Intervals Between Pulses}

Table~\ref{tab:hrampvswdta}, columns~(b) show the correlations between
the pulse amplitude hardness ratios and the intervals between the two
highest amplitude pulses in each fit.  The time intervals used are
also averaged over the two adjacent energy channels.  In all pairs of
adjacent energy channels, the two highest amplitude pulses have a
slight tendency to be closer together when the burst is harder (as
expected from cosmological redshift effects), as measured using peak
flux, but this is not statistically significant, except possibly
between channels~1 and 2.

\subsubsection{Pulse Amplitudes}

Table~\ref{tab:hrampvswdta}, columns~(c) show the correlations between
the pulse amplitude hardness ratio and the pulse amplitudes for the
highest amplitude pulse in each burst.  If the peak luminosity of the
highest amplitude pulse is a standard candle or has a narrow
distribution, the effects of cosmological redshift would introduce a
correlation between hardness ratio and amplitude.  In all pairs of
adjacent energy channels, the highest amplitude pulse has a slight
tendency to be stronger when the burst is harder, as measured using
peak flux, but this is not statistically significant (except possibly
between channels~2 and 3,) indicating that the distribution of the
above-mentioned luminosity is broad.

Table~\ref{tab:hrampfwhmamp}, columns~(b) shows the correlations
between pulse amplitude hardness ratios and pulse amplitudes within
bursts.  The pulse amplitudes are summed over the two adjacent energy
channels that the hardness ratios are taken between.  There appears to
be no statistically significant tendency for higher amplitude pulses
to have harder or softer spectra, although slightly more bursts show a
positive correlation (higher amplitude pulses are harder) than a
negative correlation (higher amplitude pulses are softer.)  This
points to a weak or negligible intrinsic correlation between these
quantities.

\subsubsection{Count Fluence Hardness Ratios}

In what follows, we carry out the same tests using the hardness ratio
measured by count fluence instead of amplitude, for bursts and pulses
within bursts.

Table~\ref{tab:hrtareavswdttarea}, columns~(a) shows the correlations
between the total burst count fluence hardness ratios and the pulse
widths for the highest amplitude pulses of the bursts.  A positive
correlation (harder bursts having shorter durations) would be expected
if pulse total energy had a narrow intrinsic distribution.  There is
no consistent or statistically significant tendency for the highest
amplitude pulse in each burst to be wider or narrower when the burst
is harder or softer, as measured using fluence.

\begin{deluxetable}{crlrlrl}
\tablecaption{Correlations Between Total Burst Count Fluence Hardness Ratio and (a) Pulse Widths, (b) Intervals Between Two Highest Amplitude Pulses, and (c) Total Burst Count Fluences for Highest Amplitude Pulse(s) in Each Burst. \label{tab:hrtareavswdttarea}}
\tablehead{
\colhead{Energy} & \multicolumn{2}{c}{(a) Width} & \multicolumn{2}{c}{(b) Interval} & \multicolumn{2}{c}{(c) Count Fluence} \\
\colhead{Channels} & \colhead{$r_{s}$} & \colhead{Prob.} & \colhead{$r_{s}$} & \colhead{Prob.} & \colhead{$r_{s}$} & \colhead{Prob.}}
\startdata
2 / 1 & 0.02 & 0.84 & -0.10 & 0.36 & -0.01 & 0.88 \\
3 / 2 & -0.22 & 0.0078 & -0.07 & 0.52 & 0.00 & 0.98 \\
4 / 3 & 0.10 & 0.42 & -0.23 & 0.19 & 0.21 & 0.084 \\
\enddata
\end{deluxetable}

Table~\ref{tab:hrareafwhmarea}, columns~(a) show the correlations
between pulse count fluence hardness ratios and pulse widths within
bursts.  In channels~1 and 2, more bursts show negative correlations
between the two quantities, \emph{i.e.} longer duration pulses tend to
have softer spectra, as measured using count fluence, and this effect,
which may be statistically significant, indicates the presence of an
intrinsic correlation.  There are no statistically significant effects
between channels~2 and 3 or between channels~3 and 4.

\begin{deluxetable}{crlrl}
\tablecaption{Correlations Between Pulse Count Fluence Hardness Ratio and (a) Pulse Width and (b) Count Fluence Within Bursts. \label{tab:hrareafwhmarea}}
\tablehead{
\colhead{Energy} & \multicolumn{2}{c}{(a) Width} & \multicolumn{2}{c}{(b) Count Fluence} \\
\colhead{Channels} & \colhead{\% Pos. Corr.} & \colhead{Binom. Prob.} & \colhead{\% Pos. Corr.} & \colhead{Binom. Prob.}}
\startdata
2 / 1 & 52.5/83 = 63\% & 0.016 & 51.5/83 = 62\% & 0.028 \\
3 / 2 & 47.5/95 = 50\% & 1.0 & 47/95 = 49\% & 0.49 \\
4 / 3 & 18.5/33 = 56\% & 0.49 & 20/33 = 66\% & 0.61 \\
All & 118.5/211 = 56\% & \nodata & 118.5/211 = 56\% & \nodata \\
\enddata
\end{deluxetable}

Table~\ref{tab:hrtareavswdttarea}, columns~(b) shows the correlations
between the total burst count fluence hardness ratios and the
intervals between the two highest amplitude pulse in each fit.  In all
pairs of adjacent energy channels, the two highest amplitude pulses
have a slight tendency to be closer together when the burst is harder
(as expected from cosmological effects,) but this is not statistically
significant.

Table~\ref{tab:hrtareavswdttarea}, columns~(c) shows the correlations
between the total burst count fluence hardness ratios and the total
burst count fluence in each fit.  There is no consistent or
statistically significant tendency for harder or softer bursts to
contain fewer or more counts.

Table~\ref{tab:hrareafwhmarea}, columns~(b) shows the correlations
between pulse count fluence hardness ratios and pulse count fluences
within bursts.  The pulse count fluences are summed over the two
adjacent energy channels that the hardness ratios are taken between.
In channels~1 and 2, more bursts show negative correlations,
\emph{i.e.}  higher fluence pulses tend to have softer spectra, but
again, this intrinsic effect appears weak, and there is no significant
effect in the other pairs of energy channels.

\emph{In summary, there seems to be little intrinsic correlation
between the spectra, as measured by hardness ratio, and other pulse
characteristics between bursts and among pulses.  There may be weak
(statistically not very significant) evidence for trends expected from
cosmological redshift effects.}

\subsection{Time Evolution of Pulse Characteristics Within Bursts}

One class of correlations between pulse characteristics within bursts
are those between the pulse peak time and other pulse characteristics.
These indicate whether certain pulse characteristics tend to evolve in
a particular way during the course of a burst.  Again, we have used
the method described in the previous section, calculating the Spearman
rank-order correlation coefficients for the individual bursts, and
testing the observed numbers of bursts with positive and negative
correlations using the binomial distribution.

\subsubsection{Pulse Asymmetry Ratios}

Table~\ref{tab:rdfwhmtmax}, columns~(a) show the number and fraction
of bursts (in each channel) where there is a negative correlation
between the pulse asymmetry ratio $\sigma_{r}/\sigma_{d}$ and peak
time, \emph{i.e.}, the pulse asymmetry decreases with time.  Fits for
which the calculated Spearman rank-order correlation coefficient was
0, indicating no correlation, were counted as half for decreasing and
half for increasing in order to calculate, using the binomial
distribution, the probability of this occurring randomly if pulse
amplitudes within bursts are equally likely to increase as to decrease
with time.  The probability was not calculated for all energy channels
combined, because fits to the same burst in different energy channels
cannot be considered independent, so the binomial distribution cannot
be used.

Pulse asymmetry ratios more often decrease than increase with time
during bursts, except in energy channel~4, which has the fewest
pulses.  This effect appears to be statistically significant in
channel~3, and possibly channels~1 and 2.  The fits to simulations
(See Table~\ref{tab:sim_rdtmax}) show no tendency for pulse
asymmetries to increase or decrease within bursts.  This indicates
that the observed tendency for pulse asymmetry ratios to decrease with
time within actual bursts does not arise from selection effects in the
pulse-fitting procedure, so that any tendency would be intrinsic to
gamma-ray bursts.

\begin{deluxetable}{crlrl}
\tablecaption{Correlations Between (a) Pulse Asymmetry Ratio and (b) Width and Pulse Peak Time Within Bursts. \label{tab:rdfwhmtmax}}
\tablehead{
\colhead{Energy} & \multicolumn{2}{c}{(a) $\sigma_{r} / \sigma_{d}$} & \multicolumn{2}{c}{(b) Width} \\
\colhead{Channel} & \colhead{\% Decreasing} & \colhead{Binom. Prob.} & \colhead{\% Decreasing} & \colhead{Binom. Prob.}}
\startdata
1 & 61.5/94 = 66\% & 0.0028 & 47.5/94 = 51\% & 0.92 \\
2 & 66.5/109 = 61\% & 0.022 & 60/109 = 55\% & 0.29 \\
3 & 81/116 = 70\% & \e{1.9}{-5} & 64/116 = 55\% & 0.27 \\
4 & 16/35 = 45\% & 0.61 & 23.5/35 = 67\% & 0.043 \\
All & 225/354 = 64\% & \nodata & 195/354 = 55\% & \nodata \\
\enddata
\end{deluxetable}

\subsubsection{Pulse Rise and Decay Times and Pulse Widths}

When we examine the evolution of the rise and decay times separately,
instead of their ratios, and the evolution of the pulse widths, we
find that there is a nearly equal and opposite trend of decreasing
rise times $\sigma_{r}$ and increasing decay times $\sigma_{d}$ as the
burst progresses.  This gives rise to the evolution of pulse asymmetry
ratios described above, although the statistical significance of the
evolution of rise times and decay time are weaker than for the pulse
asymmetry ratios.  The decrease in rise times is possibly a slightly
stronger effect than the increase in decay times.  However, the
combined effect of these two trends is that there appears to be no
statistically significant evolution of pulse widths.  (See
Table~\ref{tab:rdfwhmtmax}, columns~(b).)  This is in agreement with
the results of \cite{ramirez-ruiz:1999,ramirez-ruiz:1999b}, who found
no evidence that pulse widths increase or decrease with time when
fitting a power-law time dependence, using a small sample of complex
bursts selected from the bright, long bursts fitted by
\cite{norris:1996}.

\subsubsection{Spectra}

It has been previously reported that bursts tend to show a
hard-to-soft spectral evolution, which we can test by seeing how the
hardness ratios of individual pulses vary with time.
Table~\ref{tab:hramphrareatmax}, columns~(a) show that the pulse
amplitude hardness ratios have a slight tendency to decrease with time
during bursts between all three pairs of adjacent energy channels.
However, with the numbers of available bursts that are composed of
multiple pulses in adjacent energy channels, this tendency is
statistically insignificant.

\begin{deluxetable}{crlrl}
\tablecaption{Correlations Between (a) Pulse Amplitude and (b) Count Fluence Hardness Ratios and Pulse Peak Time Within Bursts. \label{tab:hramphrareatmax}}
\tablehead{
\colhead{Energy} & \multicolumn{2}{c}{(a) Amplitude HR} & \multicolumn{2}{c}{(b) Count Fluence HR} \\
\colhead{Channels} & \colhead{\% Decreasing} & \colhead{Binom. Prob.} & \colhead{\% Decreasing} & \colhead{Binom. Prob.}}
\startdata
2 / 1 & 46.5/83 = 56\% & 0.27 & 42.5/83 = 51\% & 0.83 \\
3 / 2 & 54.5/95 = 57\% & 0.15 & 61/95 = 64\% & 0.0056 \\
4 / 3 & 18.5/33 = 56\% & 0.48 & 13/33 = 39\% & 0.22 \\
All & 119.5/211 = 57\% & \nodata & 116.5/211 = 55\% & \nodata \\
\enddata
\end{deluxetable}

\subsubsection{Pulse Count Fluences}

When we consider the time evolution of the pulse count fluence
hardness ratios within bursts, we find no tendency for the hardness
ratio of energy channel~2 to channel~1 to increase or decrease, a
possibly significant tendency for the hardness ratio of energy
channel~3 to 2 to decrease with time, and a statistically
insignificant tendency for the hardness ratio of channel~4 to 3 to
increase with time.  (See Table~\ref{tab:hramphrareatmax},
columns~(b).)

\subsubsection{Other Pulse Characteristics}

We have conducted similar tests for other pulse characteristics and
found that none show any tendencies to increase or decrease with time
within bursts that are clearly statistically significant
\citep{lee:thesis}.  These pulse characteristics are the pulse
amplitude, the peakedness parameter $\nu$, and the pulse count
fluence; \emph{e.g.}, we find no tendency for later pulses within a
burst to be stronger or weaker, than earlier pulses.

\emph{In summary, we find no significant correlations between the peak
times of pulses in bursts and any other pulse characteristics except
possibly the pulse asymmetry ratio, so that the pulses appear to
result from random and independent emission episodes.}

\section{Discussion}
\label{sec:discuss}

Decomposing burst time profiles into a superposition of discrete
pulses gives a compact representation that appears to contain their
important features, so this seems to be a useful approach for
analyzing their characteristics.  Our pulse decomposition analysis
confirms a number of previously reported properties of gamma-ray burst
time profiles using a larger sample of bursts with generally finer
time resolution than in prior studies.  These properties include
tendencies for the individual pulses comprising bursts to have shorter
rise times than decay times; for the pulses to have shorter durations
at higher energies; and for the pulses to peak earlier at higher
energies, which is sometimes described as a hard-to-soft spectral
evolution of individual pulses.

Pulse rise times tend to decrease during the course of a burst, while
pulse decay times tend to increase.  When examining pulse widths, or
durations, these two effects nearly balance each other; the apparent
tendency for pulse widths to decrease during the course of a burst
appears to be statistically insignificant.  The ratios of pulse rise
times to decay times tend to decrease during the course of a burst.
The evolution of pulse asymmetry ratios does not arise from selection
effects in the pulse-fitting procedure, so it is most likely intrinsic
to the bursters.

No other pulse characteristics show any time evolution within bursts,
although it is possible that there is non-monotonic evolution; for
example, a pulse characteristic may tend to be greater at the
beginning and end of a burst and smaller in the middle, and the tests
used here wouldn't be sensitive to this.  In particular, it doesn't
appear that either pulse amplitudes or pulse count fluences have any
tendency to increase or decrease during the course of a burst.  Also,
later pulses in a burst don't tend to be spectrally harder or softer
than earlier pulses, although there is spectral softening
\emph{within} most pulses.  The spectra of pulses within a burst also
don't appear to be harder or softer for stronger or weaker pulses, or
for longer or shorter duration pulses.

One may therefore conclude that the pulses in a burst arise from
random and independent emission episodes such as those expected in the
internal episodic shock model rather than the external shock models
where the presence of distinguishable pulses must be attributed to
inhomogeneities in the interaction of the blast wave shock and the
clumpy interstellar medium.

When examining similar correlations between the attributes of some
characteristic pulses from burst to burst, we find some weak and
tantalizing evidence which may be due to cosmological redshift
effects.  In the accompanying paper \cite{lee:2000b} we describe the
correlation studies which can distinguish between trends due to
cosmological redshifts and intrinsic trends.

\acknowledgments

We thank Jeffrey Scargle and Jay Norris for many useful discussions.
This work was supported in part by Department of Energy contract
DE--AC03--76SF00515.

\appendix

\section{Appendix: Testing for Selection Effects}

There are a number of ways in which the pulse-fitting procedure may
introduce selection effects into correlations between pulse
characteristics.  One is that the errors in the different fitted pulse
parameters may be correlated.  Another is that the pulse-fitting
procedure may miss some pulses by not identifying them above the
background noise.  Still another cause of selection effects is that
overlapping pulses may be identified as a single broader pulse.

In order to determine the degree of importance of these selection
effects, we have generated a sample of artificial burst time profiles
using the pulse model with randomly generated but known pulse
parameters, fitted the simulated bursts using the same procedure used
for actual burst data, and compared the simulated and fitted pulse
characteristics~\citep{lee:thesis}.

\subsection{Numbers of Bursts and Pulses}

\begin{figure}\plotone{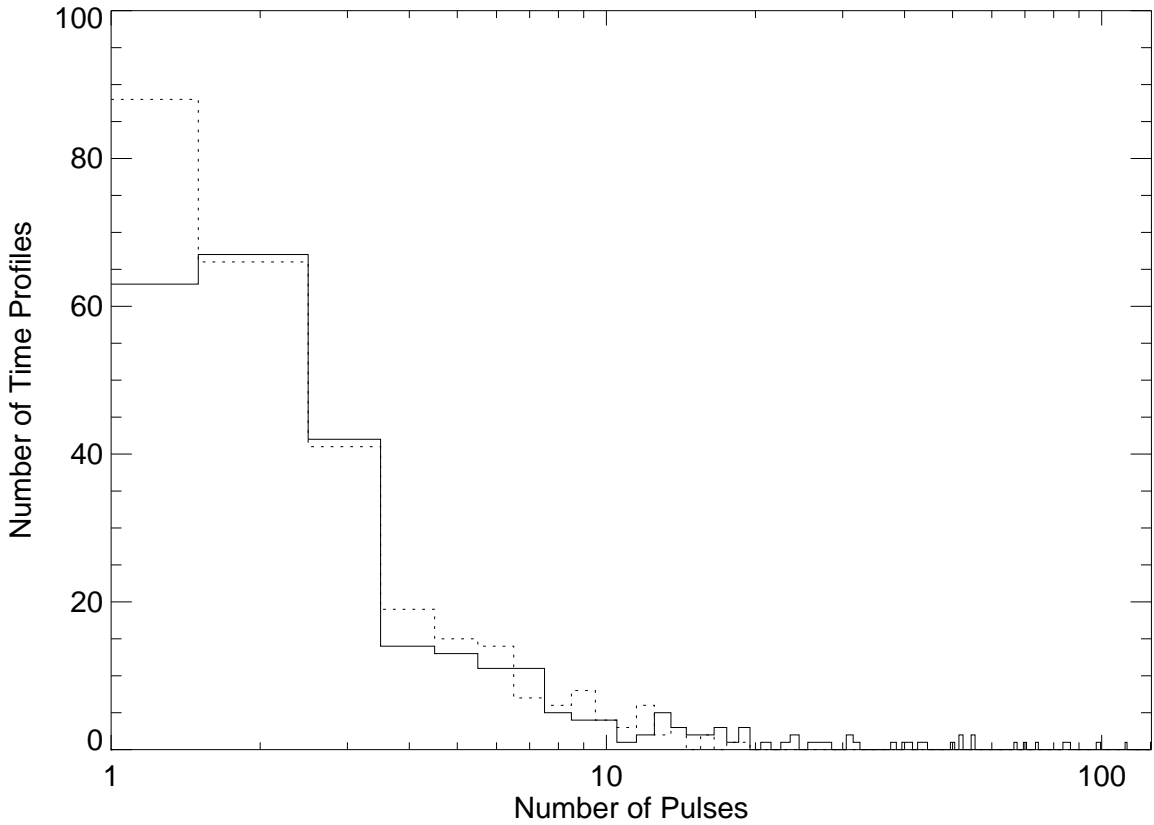}\caption{Distribution of number of pulses in initial simulations (solid histogram) and in the results of the fits to the simulated data (dashed histogram).  \label{sim_npulse}}
\end{figure}

\begin{figure}\plotone{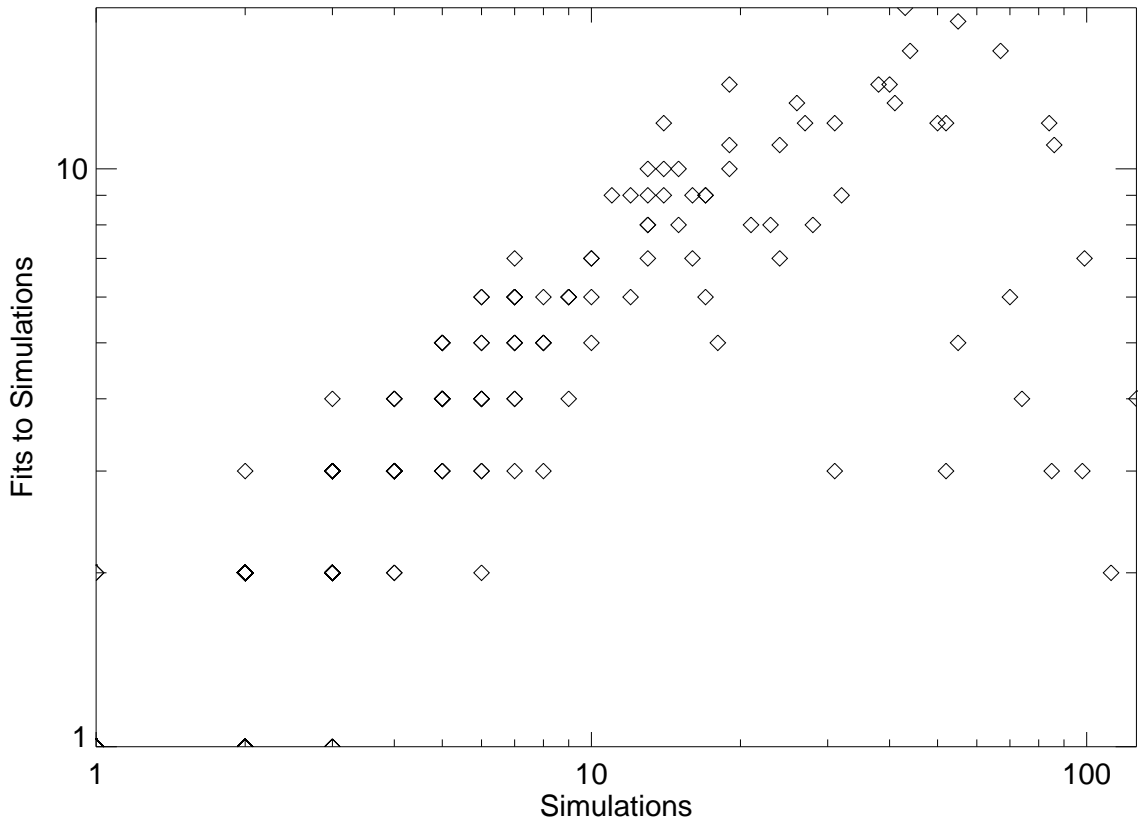}\caption{Number of pulses obtained from fits to the simulations versus number in the initial simulations.  Note that the
selection effect of the fitting procedure is more pronounced in bursts
with larger numbers of pulses.  \label{fitvssim}}
\end{figure}

\begin{figure}\plotone{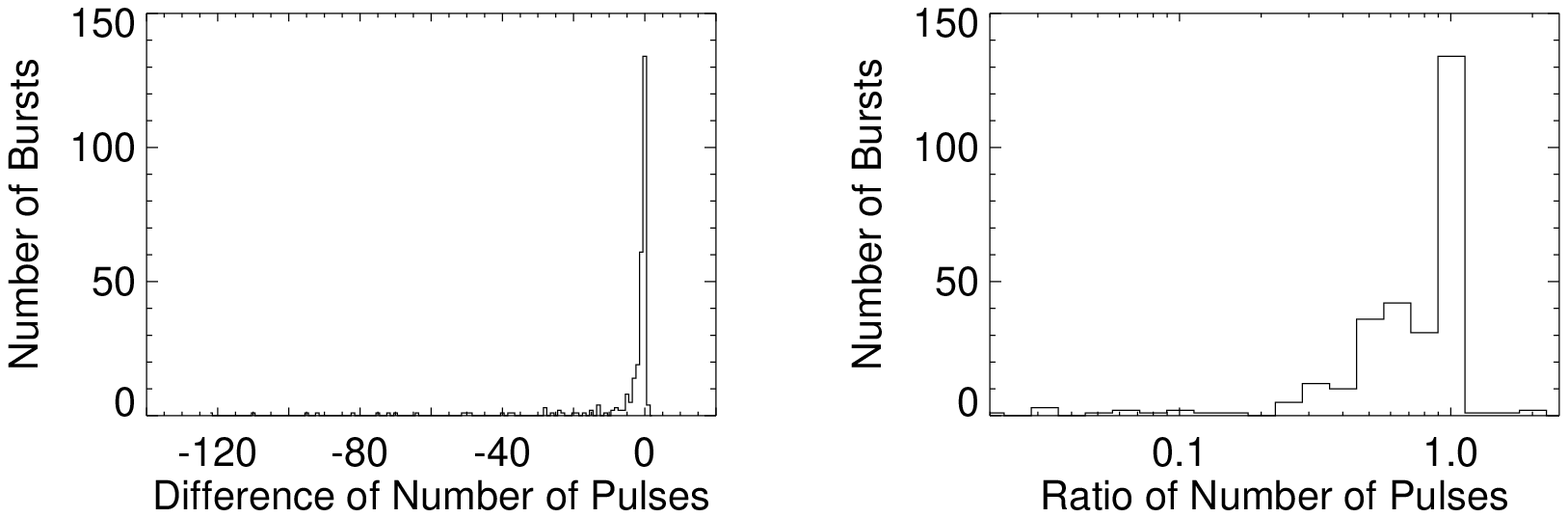}\caption{(Left) Distribution of differences between
numbers of pulses from initial simulated bursts to fits to simulated
bursts.  A small number of bursts have many fewer pulses in the fits
to simulations than in the initial simulations.  (Right) Distribution
of ratios of numbers of pulses from initial simulated bursts to fits
to simulated bursts.  \label{sim_rdnp}}
\end{figure}

A total of 286 bursts were generated, with only one energy channel for
each burst.  For many of these, the limit of $2^{20}$ counts was
reached before the 240 second limit, which almost never occurred in
the actual BATSE TTS data.  These simulated bursts contained a total
of 2671 pulses that had peak times before the limits of $2^{20}$
counts and 240 seconds, while the fits to the simulated bursts
contained a total of only 1029 pulses.  Of these, 223 of the simulated
bursts and 198 of the fits to the simulations contained more than one
pulse.  (See Figure~\ref{sim_npulse} and Table~\ref{tab:sim_npulse}.)
Note that in the fits to actual BATSE data, the largest number of fits
containing more than one pulse was 116 for energy channel~3, so that
the simulated data set is larger.  Figure~\ref{fitvssim} shows the
number of pulses fitted versus the number of pulses originally
generated for each simulated bursts.  It shows that the greatest
differences between the the fitted and the simulated numbers of pulses
tend to occur in the most complex bursts.  Figure~\ref{sim_rdnp}
compares the numbers of pulses per fit between the simulations and the
fits to simulations.  Most (54\%) of the fits to simulated bursts
contain fewer pulses than the initial simulations, and for nearly all
of the remaining simulated bursts, the number of pulses are the same
for the initial simulations and the fits to simulations.  The fits to
simulations have a mean of 15 fewer pulses than the initial
simulations, and a median of 1 fewer pulse.  The fits to simulations
have a geometric mean of 0.63 times as many pulses as the initial
simulations, and a mean of 0.80 times as many pulses.

\begin{deluxetable}{crrrr}
\tablecaption{Characteristics of Distribution of Number of Pulses in Simulations. \label{tab:sim_npulse}}
\tablehead{
\colhead{} & \colhead{Median No.} & \colhead{Mean No.} & \colhead{Max. No.} & \colhead{\% Single} \\
\colhead{} & \colhead{of Pulses} & \colhead{of Pulses} & \colhead{of Pulses} & \colhead{Pulse}}
\startdata
Simulation & 3 & 9.3 & 126 & 62/286 = 22\% \\
Fit to Sim. & 2 & 3.6 & 19 & 88/286 = 31\% \\
\enddata
\end{deluxetable}

\subsection{Brightness Measures of Simulated Pulses}
%\subsection{Pulse Amplitudes}

\begin{figure}\plotone{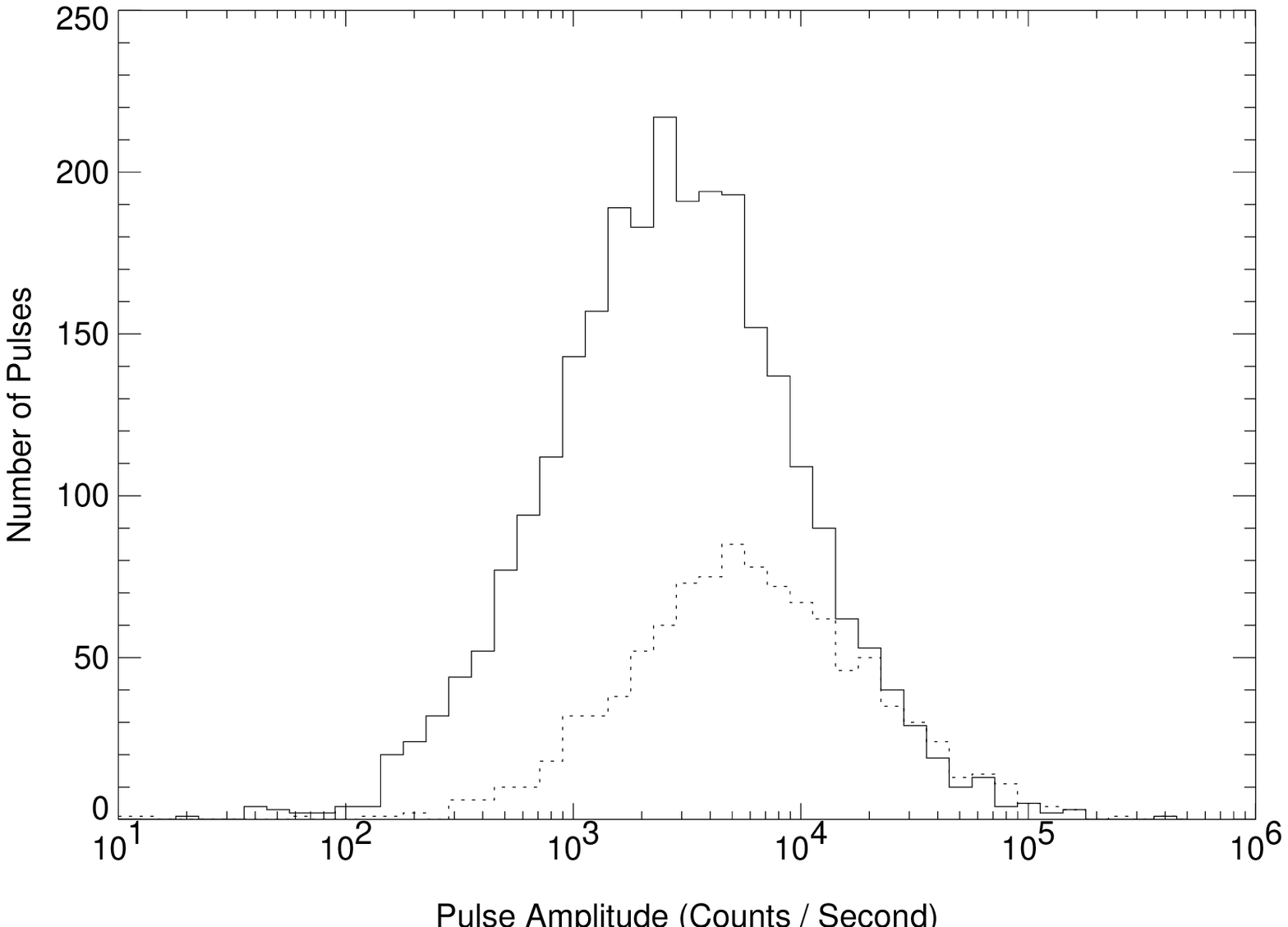}\caption{Distribution of pulse amplitudes for all pulses from all bursts in initial simulations (solid histogram) and in the results of the fits to the simulated data (dashed histogram). \label{sim_amp}}
\end{figure}

Figure~\ref{sim_amp} shows the distribution of pulse amplitudes in the
original simulations and in the fits to simulations.  It shows that
the fitting procedure has a strong tendency to miss low amplitude
pulses.  However, if we compare this with Figure~\ref{amp}, we see
that in the fits to actual BATSE bursts, the fitting procedure found
pulses with considerably lower amplitudes than it found in the fits to
simulated bursts.

\begin{figure}\plotone{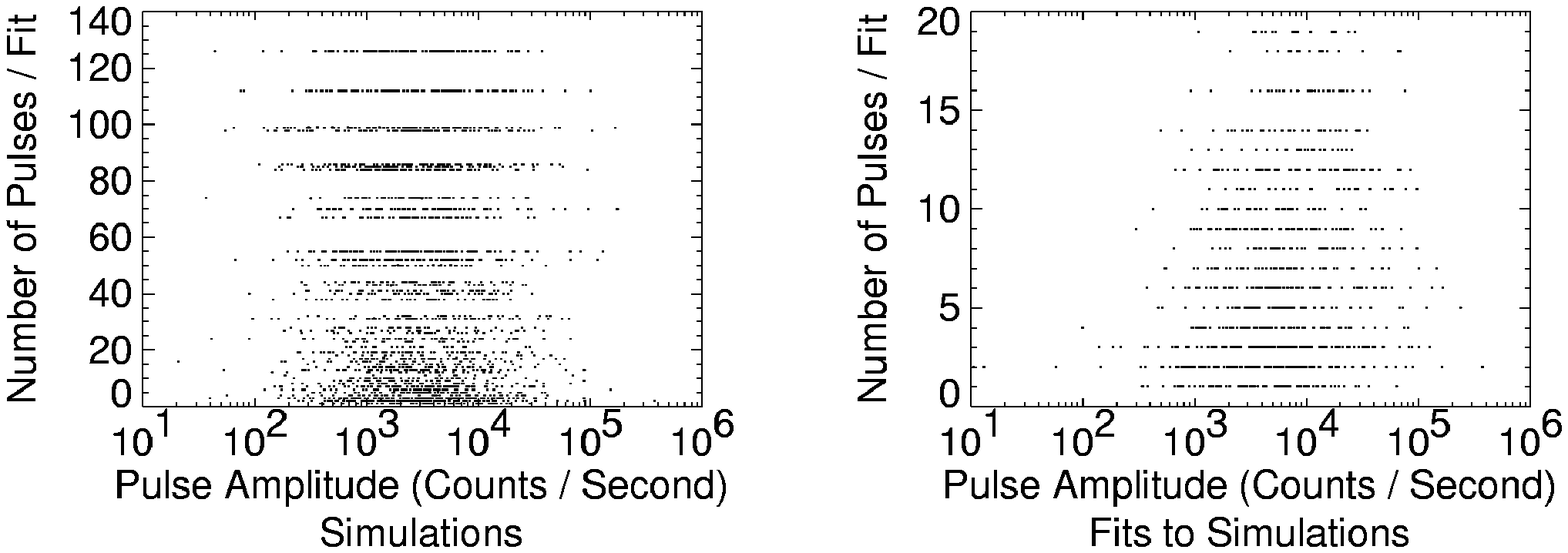}\caption{Number of pulses per fit versus pulse amplitudes of all pulses. \label{sim_npvsamp}}
\end{figure}

Figure~\ref{sim_npvsamp} shows the number of pulses in each burst
plotted against the amplitudes of all of the pulses comprising each
fit.  In the simulations, there are no correlations between pulse
amplitudes and the number of pulses in the time profile, because the
pulse amplitudes were generated independently of the number of pulses
in each burst.  In the fits to the simulations, pulse amplitudes tend
to be higher in bursts containing more pulses.  This must result from
the selection effect discussed in Section~\ref{sec:amp}; it is easier
to identify more pulses when they are stronger.
Table~\ref{tab:sim_npvsafw}, columns~(a) shows that even though there
is no correlation between pulse amplitudes and the number of pulses
for the initial simulated data, the fitting procedure introduces a
strong positive correlation between these quantities; the tendency to
miss low amplitude pulses is greater in more complex bursts.

\begin{deluxetable}{crlrlrl}
\tablecaption{Correlation Between Number of Pulses per Burst and (a) Amplitudes, (b) Count Fluences, and (c) Widths of All Pulses in Simulated Bursts. \label{tab:sim_npvsafw}}
\tablehead{
\colhead{} & \multicolumn{2}{c}{(a) Amplitude} & \multicolumn{2}{c}{(b) Count Fluence} & \multicolumn{2}{c}{(c) Width} \\
\colhead{} & \colhead{$r_{s}$} & \colhead{Prob.} & \colhead{$r_{s}$} & \colhead{Prob.} & \colhead{$r_{s}$} & \colhead{Prob.}}
\startdata
Simulation & -0.04 & 0.042 & -0.02 & 0.31 & 0.01 & 0.76 \\
Fit to Sims. & 0.22 & \e{7.1}{-13} & 0.11 & 0.00068 & -0.03 & 0.36 \\
\enddata
\end{deluxetable}

%\subsection{Count Fluences}

\begin{figure}\plotone{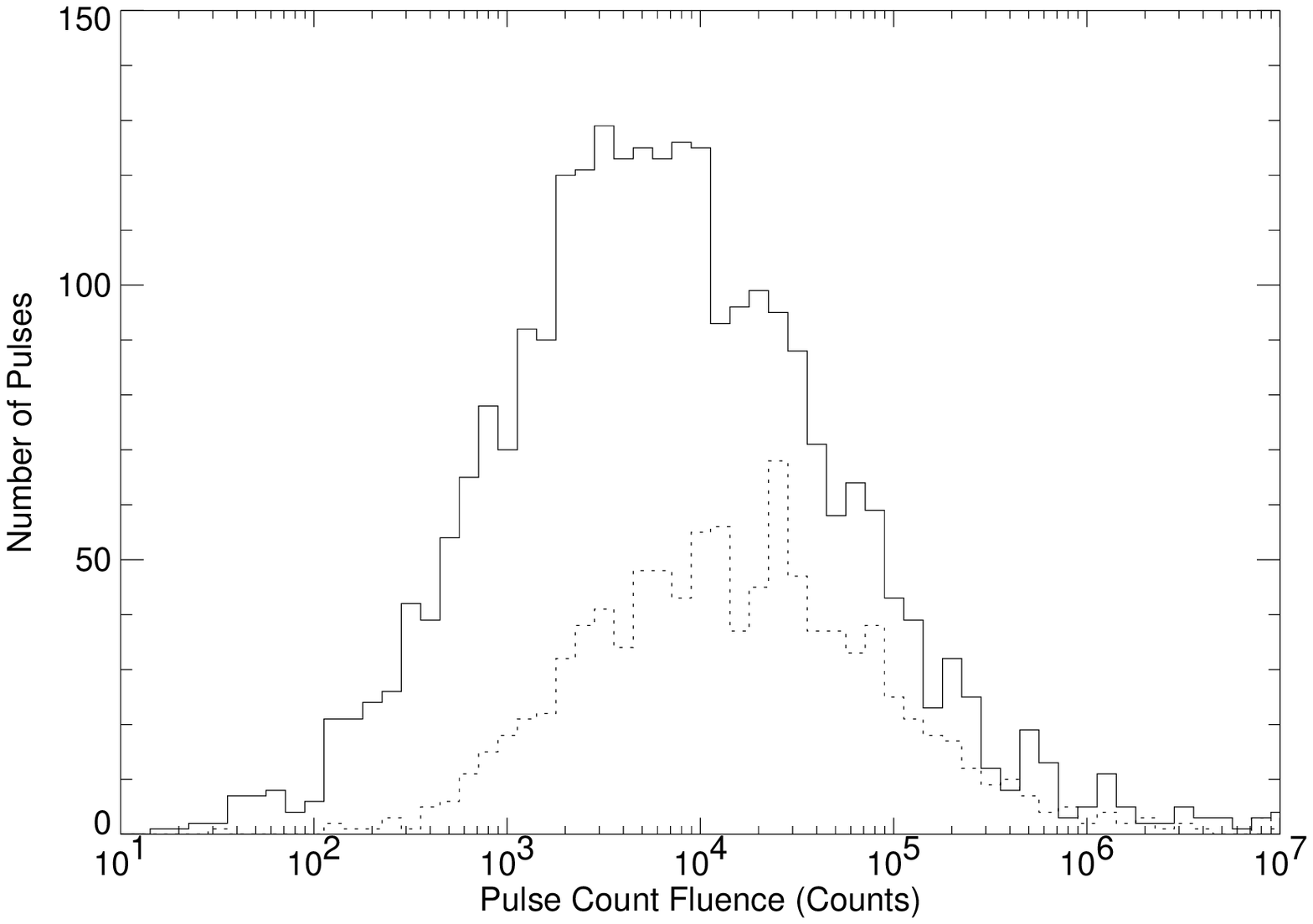}\caption{Distribution of pulse count fluences for all pulses from all bursts in initial simulations (solid histogram) and in the results of the fits to the simulated data (dashed histogram). \label{sim_area}}
\end{figure}

Figure~\ref{sim_area} shows the distribution of pulse count fluences
in the original simulations and in the fits to simulations.  It shows
that the fitting procedure has a strong tendency to miss pulses with
low count fluences, similar to what we have seen for low amplitude
pulses.

\begin{figure}\plotone{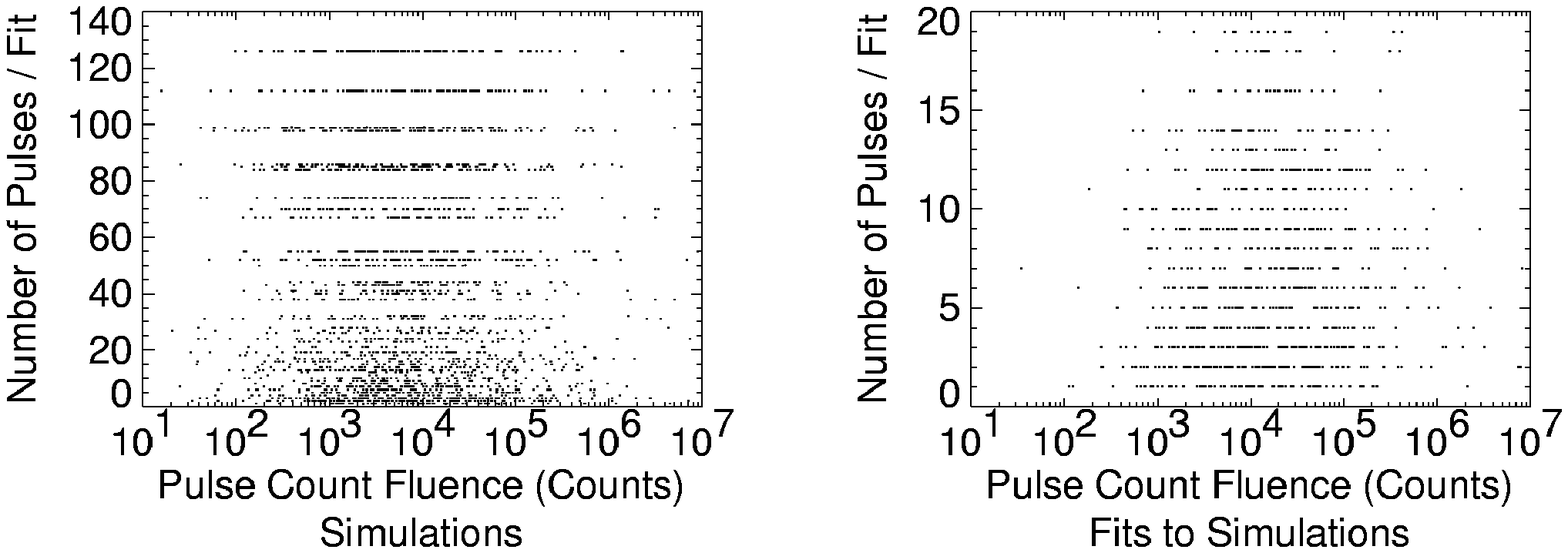}\caption{Number of pulses per burst versus count fluences of all pulses. \label{sim_npvsarea}}
\end{figure}

Figure~\ref{sim_npvsarea} and Table~\ref{tab:sim_npvsafw}, columns~(b)
compare the number of pulses in each time profile with the count
fluences of the individual pulses.  They show no tendency for pulses
to contain fewer or more counts in bursts with more pulses, in either
the initial simulations (by design) or in the fits to simulations.
Unlike pulse amplitudes, the tendency to miss low count fluence pulses
appears to be independent of burst complexity.  This differs from the
results seen in the fits to actual bursts, where bursts containing
more pulses tended to have pulses with lower count fluences.  (See
Figure~\ref{npvsarea} and Table~\ref{tab:npvsafw}, columns~(b).)  This
may explain why the $2^{20}$ count limit for the TTS data was
frequently reached before the 240 second time limit in the simulated
bursts, but rarely in the actual bursts; the total count fluence
increases linearly with the number of pulses in the simulated bursts,
but less rapidly in the actual BATSE bursts.

\subsection{Pulse Widths}

\begin{figure}\plotone{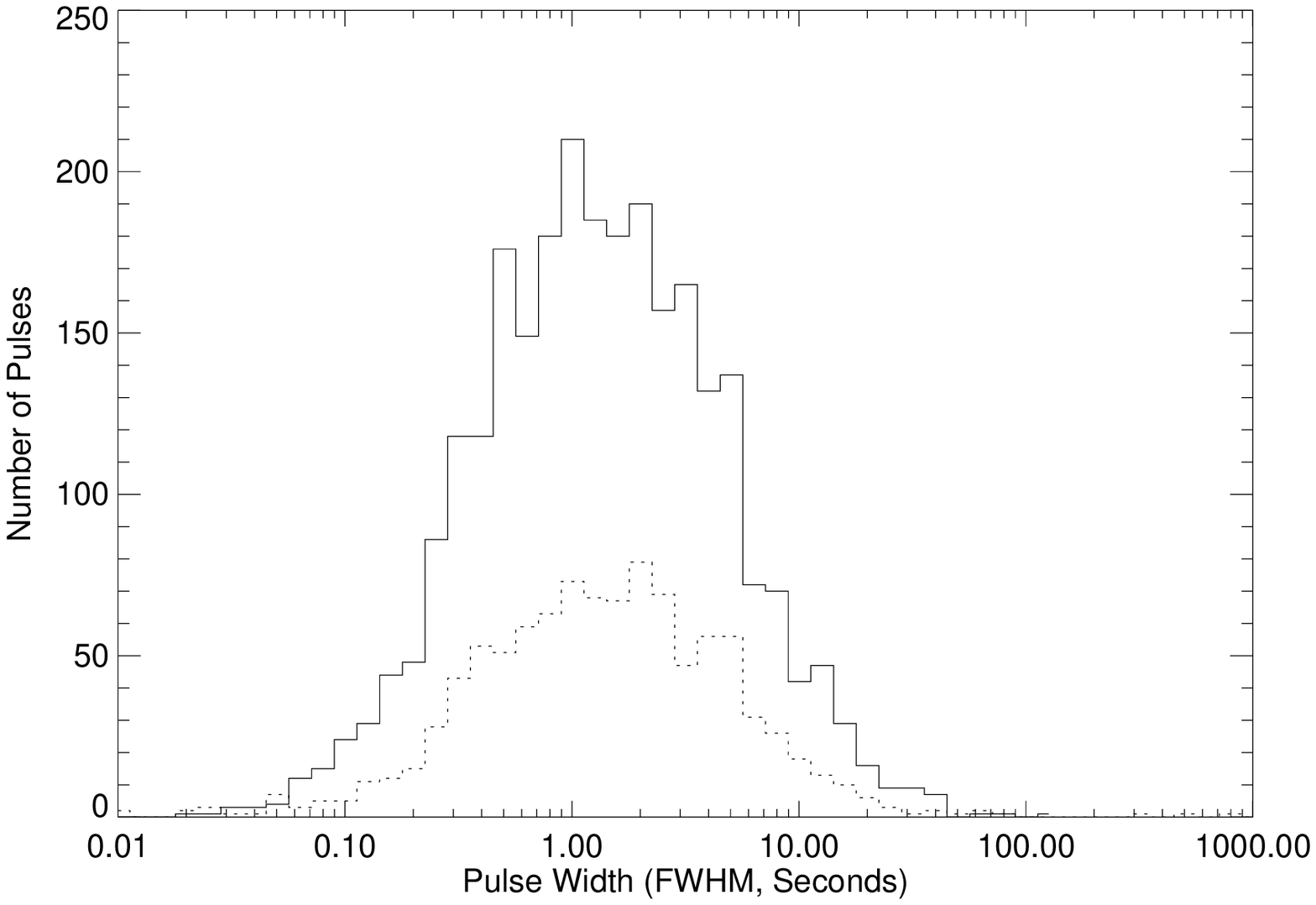}\caption{Distribution of pulse widths for all pulses from all bursts in initial simulations (solid histogram) and in the results of the fits to the simulated data (dashed histogram). \label{sim_fwhm}}
\end{figure}

Figure~\ref{sim_fwhm} shows the distribution of pulse widths in the
original simulations and in the fits to simulations.  The pulses in
the fits to simulations tend to be slightly longer in duration than in
the original simulations, but applying the Kolmogorov-Smirnov test to
the two distributions show that they are not significantly different;
the probability that they are the same distribution is 0.39.  This
agrees with what we have seen in Figures~\ref{sim_amp} and
\ref{sim_area}, that the selection effects of the fitting procedure
for pulse amplitudes and for pulse count fluences are similar.

\begin{figure}\plotone{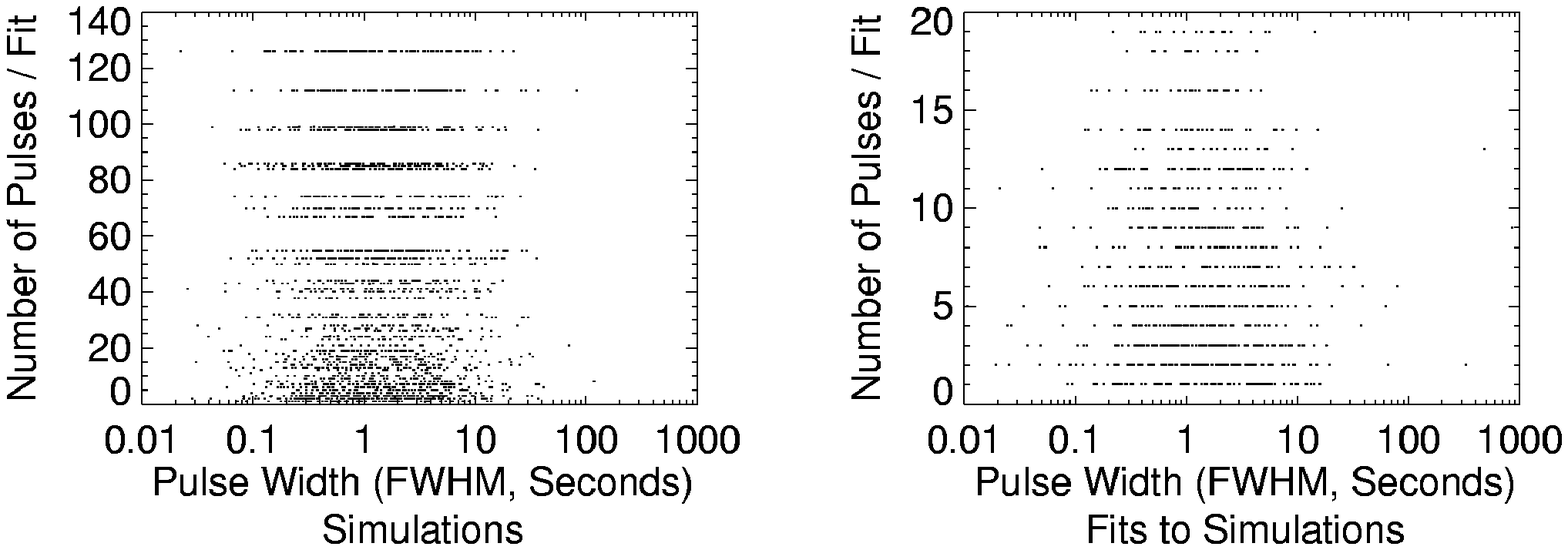}\caption{Number of pulses per burst versus widths of all pulses. \label{sim_npvsfwhm}}
\end{figure}

Figure~\ref{sim_npvsfwhm} and Table~\ref{tab:sim_npvsafw}, columns~(c)
compare the number of pulses in each time profile with the widths of
the individual pulses.  They show no tendency for pulses to be wider
or narrower in bursts with more pulses, in either the simulations or
the fits to the simulations.

\subsection{Time Evolution of Pulse Characteristics Within Bursts}

In the fits to actual BATSE data, it was found that pulse asymmetry
ratios tended to decrease over the course of a burst.  (See
Table~\ref{tab:rdfwhmtmax}.)  Table~\ref{tab:sim_rdtmax} shows the
correlations between pulse asymmetry ratio and peak times within
bursts for the simulations and the fits to simulations.  It shows no
tendency for positive or negative correlations in either the
simulations or the fits to simulations.

\begin{deluxetable}{crl}
\tablecaption{Correlations Between Pulse Asymmetry Ratio and Pulse Peak Time Within Simulated Bursts. \label{tab:sim_rdtmax}}
\tablehead{
\colhead{} & \colhead{\% Decreasing} & \colhead{Binom. Prob.}}
\startdata
Simulation & 92.5/223 = 41\% & 0.016 \\
Fit to Sim. & 103/198 = 52\% & 0.57 \\
\enddata
\end{deluxetable}

{
%\bibliographystyle{abbrvnat}
%\bibstyle@aa
%\bibliographystyle{apj}
%\bibliography{apj-jour,alee}

}

\end{document}